\documentclass[final,3p,times,twocolumn,numbers,sort,compress]{elsarticle}

\usepackage{graphicx}
\usepackage{amsmath,physics,bm,float}
\usepackage{soul}
\usepackage[colorlinks=true,citecolor=blue,urlcolor=blue,linkcolor=blue]{hyperref}
\usepackage{amssymb}
\usepackage{natbib}
\usepackage{upgreek}
\usepackage{mathtools}

\DeclareMathOperator*{\argmin}{argmin}

\DeclareMathOperator*{\ourmod}{\,mod\,}
\newcommand{\rms}{\rm\scriptscriptstyle}
\newcommand{\nb}{n_{\rm b}}
\newcommand{\mb}{\lceil\nb\sigma\rceil}
\newcommand{\nc}{n_{\rm c}}

\newcommand{\nbc}{n_{\rm bc}}
\newcommand{\ncb}{\nc+\nb}
\newcommand{\mbc}{\lceil (\nb+\nc)\sigma\rceil}

\newcommand{\fc}{f_{\rm c}}

\usepackage[mathlines]{lineno}
\let\oldequation\equation\let\oldendequation\endequation
\renewenvironment{equation}{\linenomathNonumbers\oldequation}{\oldendequation\endlinenomath}
\let\oldalign\align\let\oldendalign\endalign
\renewenvironment{align}{\linenomathNonumbers\oldalign}{\oldendalign\endlinenomath}

\journal{Chaos, Solitons \& Fractals}

\begin{document}

\begin{frontmatter}

\title{Solitary cluster waves in periodic potentials: Formation, propagation, and soliton-mediated particle transport}

\author[inst1,inst2]{Alexander P.\ Antonov}
		
\affiliation[inst1]{organization={Universität Osnabrück, 
Institut für Physik,
Fachbereich Mathematik/Informatik/Physik},
addressline={Barbarastraße 7},
postcode={D-49076},
city={Osnabrück},
country={Germany}}

\affiliation[inst2]{organization={Institut für Theoretische Physik II:
Weiche Materie, 
Heinrich-Heine-Universität Düsseldorf},\\
addressline={Universitätsstraße 1,},
postcode={D-40225},
city={Düsseldorf},
country={Germany}}

\affiliation[inst3]{organization={Charles University, 
Faculty of Mathematics and Physics, 
Department of Macromolecular Physics},
addressline={V Holešovičkách 2},
postcode={\\ CZ-18000},
city={Prague},
country={Czech Republic}}

\author[inst3]{Artem Ryabov}
\author[inst1]{Philipp Maass}

\ead{maass@uos.de}
		
\date{March 27, 2024, Revised May 24, 2024}

\begin{abstract}
Transport processes in crowded periodic structures are often mediated by cooperative movements of particles 
forming clusters.
Recent theoretical and experimental studies of driven Brownian motion of hard spheres showed that
cluster-mediated transport in one-dimensional periodic potentials can proceed in form of solitary waves. 
We here give a comprehensive description of these solitons. 
Fundamental for our analysis is a static presoliton state, 
which is formed by a periodic arrangement of basic stable clusters. Their size follows from a
geometric principle of minimum free space. Adding one particle to the presoliton state gives rise to solitons. 
We derive the minimal number of particles needed for soliton formation, 
number of solitons at larger particle numbers, soliton velocities and soliton-mediated particle currents. 
Incomplete relaxations of the basic clusters are responsible
for an effective repulsive soliton-soliton interaction seen in measurements.
A dynamical phase transition is predicted to occur in current-density relations at low temperatures.
Our results provide a theoretical basis for describing experiments on cluster-mediated 
particle transport in periodic potentials.

\end{abstract}

\end{frontmatter}
	
\section{Introduction}
Particle transport in densely populated periodic structures frequently proceeds by 
cooperative movements of particle assemblies.
Examples are surface crowdions in copper adatom diffusion on a strained surface \cite{Xiao/etal:2003},
and interstitial crowdions in crystals \cite{Paneth:1950, Zepeda-Ruiz/etal:2004, Korznikova/etal:2022}.
These crowdions are low-temperature configurations of interstitial atoms densely packed in one direction,
which can extend over many
lattice constants. Analogous clusterings of vacancies can occur, which are referred to
as anti-crowdions \cite{Landau/etal:1993} or voidions \cite{Matsukawa/Zinkle:2007, vanderMeer/etal:2018}.
A further example are defect motions in
colloidal monolayers, which are completely filled wells of a two-dimensional periodic or quasi-periodic optical potential 
\cite{Korda/etal:2002, Bohlein/etal:2012, Bohlein/Bechinger:2012, Vanossi/etal:2012, Brazda/etal:2018, Vanossi/etal:2020}.
Under time-dependent forcing,  cluster-mediated transport was seen
for vortices in a nanostructured superconducting film driven by an oscillating electric current \cite{deSouzaSilva/etal:2006}, paramagnetic colloids above a magnetic bubble lattice driven by a rotating magnetic field \cite{Tierno/Fischer:2014, Stoop/etal:2020, Lips/etal:2021}, and polystyrene particles 
driven by an oscillating periodic optical potential \cite{Juniper/etal:2015}.

In various of these studies \cite{Bohlein/etal:2012, Vanossi/Tosatti:2012, Vanossi/etal:2012, Brazda/etal:2018, vanderMeer/etal:2018, Vanossi/etal:2020}, 
aspects of the cooperative particle dynamics
could be successfully interpreted by resorting to the physics of the Frenkel-Kontorova (FK) model \cite{Braun/Kivshar:2004}. This model
allows for the formation of solitary waves.
When charged colloidal particles in periodic potentials are sliding under a viscous drag force, 
double occupied and vacant wells in chains of single-occupied wells act as kinks (local compression) 
and antikinks (local expansion) \cite{Bohlein/etal:2012} as in the FK model.
Similarity to the dynamics in the FK model was further demonstrated by 
Newtonian dynamics simulations of these experiments \cite{Vanossi/etal:2012}. 
In simulations of defective crystals,
particle displacements within extended defects created by a vacancy (voidions) or interstitial (crowdions) could be described
by solutions of the sine-Gordon equation, which corresponds to a continuum limit of the FK model. 

%%%%%%%%%%%%%%%%%%%%%%%%%%%%%%%%%%%%%%%%%%%%%%%
\begin{figure*}[t!]
\includegraphics[width=\textwidth]{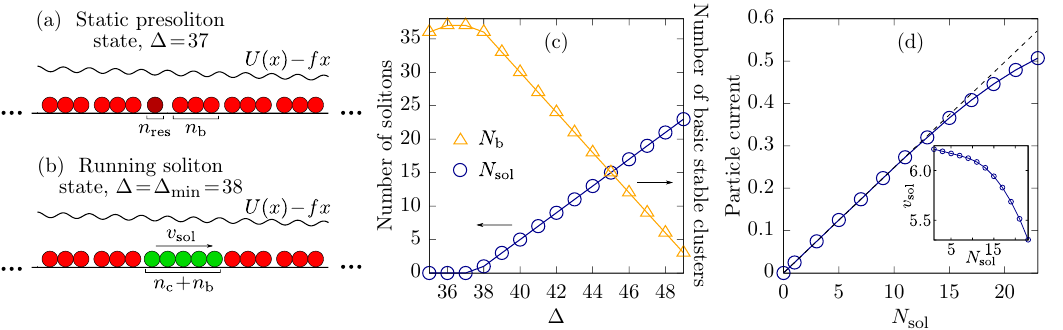}
\caption{Illustration of mechanically stable presoliton state, running state carrying solitons,
dynamical phase transition between presoliton and running state, 
and behavior of soliton-mediated particle currents.
In (a), the presoliton state in the tilted periodic potential $U(x)-fx$ 
is depicted. It is formed by a periodic sequence of basic mechanically stable clusters of size $\nb=3$ and 
one residual particle ($n_{\rm res}=1$).
In (b), a particle configuration in the running state is displayed, which forms after adding one particle 
to the presoliton state. 
One soliton cluster with five particles can be identified (spheres marked in green). 
The remaining (spheres marked in red) 
form basic $\nb$-clusters.
The configurations in (a) and (b) represent a small section of a system with $L=75$ potential wells.
How the numbers $N_{\rm sol}$ and $N_{\rm b}$ of solitons and basic stable clusters vary with
the overfilling $\Delta$ is shown in (c). $N_{\rm sol}$ jumps from zero to one
in the transition between presoliton and running state, and increases linearly with slope two 
upon further increase of $\Delta$ (two further solitons per added particle).
Part (d) shows the particle current mediated by the solitons as a function of $N_{\rm sol}$. 
For large $N_{\rm sol}$, it is not proportional to $N_{\rm sol}$
due to an effective repulsive interaction between solitons. 
This effective interaction is reflecting a slowing down of mean soliton velocities with increasing $N_{\rm sol}$, which is 
shown in the inset of (d). Results in (a)-(d) are obtained from simulations based on Eq.~\eqref{eq:langevin-zero-noise} 
with $U(x)$ from Eq.~\eqref{eq:cosine-potential},
hard sphere diameter $\sigma=0.6$, and drag force $f=0.1\fc$ [$U_0=1$, $\lambda=1$, $\mu=1$, and
$\fc$ is the critical force for overtilting, see Eq.~\eqref{eq:fc-single}].
They agree with analytical results derived in Secs.~\ref{sec:presoliton-states}-\ref{sec:particle_currents}.}
\label{fig:illustration_open_problems}
\end{figure*}
%%%%%%%%%%%%%%%%%%%%%%%%%%%%%%%%%%%%%%%%%%%%%%%

Different solitary waves of particle clusters were 
predicted to occur
in overdamped Brownian motion of 
hard spheres through one-dimensional periodic potentials \cite{Antonov/etal:2022a}.
Particle currents mediated by these solitons at low temperature showed pronounced peaks at certain magic
hard-sphere diameters.
This was found when the number $N$ of particles exceeds the number $L$ of potential wells by one.

Recently, the novel solitary cluster waves were observed in experiments \cite{Cereceda-Lopez/etal:2023}.
They uncovered new fundamental features of the soliton dynamics: the formation of multiple cluster waves when the overfilling
\begin{equation}
\Delta=N-L
\end{equation}
of the potential wells is larger than one, 
and indications of a repulsive soliton-soliton interaction in the multi-soliton states.

These findings await an explanation. 
Generally, a number of questions arises:

\begin{list}{$\bullet$}{\setlength{\leftmargin}{1.2em}\setlength{\rightmargin}{0em}
\setlength{\itemsep}{0ex}\setlength{\topsep}{0ex}}

\item Will solitary cluster waves appear for all particle diameters, if the overfilling $\Delta$ of potential wells is sufficiently high?

\item Is there a minimal overfilling $\Delta_{\rm min}$ needed to create cluster solitons?

\item How does the number of solitons vary with the overfilling $\Delta$?

\item What is the nature of the experimentally observed interaction between cluster solitons?

\item How are particle currents related to solitary cluster waves?

\end{list}
We will tackle these questions in this study. Our answers provide a basic understanding of cluster-mediated particle transport in periodic potentials.

Figure~\ref{fig:illustration_open_problems}
gives an overview of key physical mechanisms and concepts that we will discuss. 
Fundamental is the characterization of the presoliton state in Fig.~\ref{fig:illustration_open_problems}(a), which shows
a  periodic arrangement of mechanically stable particle clusters in a tilted periodic energy landscape $U(x)-fx$, where
$f$ is a drag force. 
The presoliton state arises when increasing the particle number until a further addition of a particle 
leads to a running state carrying one or more solitons, see
Fig.~\ref{fig:illustration_open_problems}(b). In the example shown 
in the figure, the presoliton is reached for overfilling $\Delta=37$ 
and one soliton appears after adding one particle, i.e.\ at the minimal overfilling
$\Delta_{\rm min}=38$ for soliton formation. Movie~1 of the Supplemental Material (SM) \cite{suppl-solitary-cluster-waves}
shows the soliton motion for the single-soliton state.

In the dynamical phase transition from the  
presoliton to the running state, the number $N_{\rm sol}$ of solitons jumps
from $N_{\rm sol}=0$ for $\Delta<\Delta_{\rm min}$
to $N_{\rm sol}^{\rm min}=1$ at  $\Delta=\Delta_{\rm min}$, see Fig.~\ref{fig:illustration_open_problems}(c).
For $\Delta>\Delta_{\rm min}$,
$N_{\rm sol}$ increases linearly with $\Delta$ with slope two, i.e.\ two further solitons are created
per added particle.
The soliton-mediated particle current $j_{\rm st}$ in the stationary state in Fig.~\ref{fig:illustration_open_problems}(d)
is proportional to the soliton number for small $N_{\rm sol}$, but the increase with
$N_{\rm sol}$ becomes sublinear for large $N_{\rm sol}$. This is a consequence of a slowing-down of the
soliton's velocity with $N_{\rm sol}$, see the inset of Fig.~\ref{fig:illustration_open_problems}(d).
The slowing down corresponds to an effective repulsive soliton-soliton interaction.

\section{Solitary cluster waves}
In general, solitons are waves that travel without dispersion and maintain their shape \cite{Ablowitz:1991, Braun/Kivshar:2004, Malomed/etal:2020}.
They were modeled and observed across various applications, including matter waves in Bose-Einstein condensates \cite{Strecker/etal:2002, Carr/Brand:2004}, optical waves in Kerr frequency combs \cite{Karpov/etal:2016, Kippenberg/etal:2018}, Alfv\'{e}n magnetic waves in plasmas \cite{Deconinck/etal:1993, Xu/etal:2008},
compression waves in damped Newtonian dynamics of Toda lattice chains subject to time-periodic driving \cite{Malomed:1992, Kuusela:1992, Kuusela/etal:1993, Hietarinta/etal:1995}, waves of stable defects in periodically inhomogeneous long Josephson junctions called supersolitons \cite{Malomed/etal:1990} or superfluxons  \cite{Malomed:1990}, as well as waves in large-scale phenomena such as tsunamis \cite{Madsen/etal:2008, Constantin/Henry:2009} and tidal bores \cite{Stanton/Ostrovsky:1998}.

Here we consider particles dragged by a constant force $f$ 
across a sinusoidal potential, 
for which first experimental observations of the solitary waves
were reported in Ref.~\cite{Cereceda-Lopez/etal:2023}.

The particles perform an overdamped Brownian motion, where effects of inertia are negligible.
This motion is described by the Langevin equations \cite{Risken:1985}
\begin{equation}
\frac{\dd x_i}{\dd t} = \mu\left[f - U'(x_i)\right] + \sqrt{2D}\,\xi_i(t)\,,
\label{eq:langevin}
\end{equation}
where $x_i(t)$,  $i=1,\ldots,N$, are the particle positions,
$\mu$ is the bare mobility, $D = \mu k_{\rms{B}}T$ is the diffusion coefficient, $k_{\rms{B}}T$ is the thermal energy, and 
$\xi_i(t)$ are Gaussian white noise processes with zero mean and correlation functions 
$\langle \xi_i(t) \xi_j(t') \rangle = \delta_{ij}\delta(t'-t)$. 
The hard-sphere interaction constrains the particle distances to 
\begin{equation}
|x_j-x_i| \ge \sigma\,,
\end{equation}
where $\sigma$ is the particle diameter.
The periodic potential $U(x)$ is
\begin{equation}
U(x) = \frac{U_0}{2}\cos\left(\frac{2\pi x}{\lambda}\right)\,,
\label{eq:cosine-potential}
\end{equation}
where $\lambda$ is the wavelength and $U_0\gg k_{\rms{B}}T$ is the barrier 
between neighboring potential wells.

With increasing drag force $f$,
the barriers of the tilted potential $U(x)-fx$ become
smaller. They disappear at the critical force 
\begin{equation}
\fc=\frac{\pi U_0}{\lambda}
\label{eq:fc-single}
\end{equation}
for overtilting.

The dynamics of the particles is constrained to a finite interval of size $L$ with  
periodic boundary conditions, where $L$ is an integer multiple of $\lambda$.

In our previous work \cite{Antonov/etal:2022a}, we treated a special case of particle diameters in the range
$0.4\,\lambda\le\sigma<\lambda$ for overfilling $\Delta=N-L=1$, i.e.\
when the number $N$ of particles exceeds the number $L$ of potential wells by one.
For high barriers $U_0\gg k_{\rms B} T$ of the periodic potential and weak
drag force $f\ll\fc$, we found that particle currents as a function of $\sigma$ exhibit peaks at
magic particle diameters $\sigma_n/\lambda=(n-1)/n$, $n=2,3,\ldots$
Solitary cluster waves, which manifest themselves
as periodic sequences of cluster movements, are responsible for this striking behavior.
Their properties could be derived by considering 
the limit of zero noise, where Eqs.~\eqref{eq:langevin} are
\begin{equation}
\frac{\dd x_i}{\dd t} = \mu\left[f - U'(x_i) \right]\,.
\label{eq:langevin-zero-noise}
\end{equation}

The situation becomes much more complex when considering arbitrary overfillings $\Delta$ and
particle diameters $\sigma$.
In this study we tackle this problem based on Eqs.~\eqref{eq:langevin-zero-noise}
and derive general conditions for the appearance of solitons and describe 
their properties.

In Sec.~\ref{sec:clusters} we first give basic features of particle clusters in a tilted periodic potential. In
Sec.~\ref{sec:presoliton-states} we discuss 
presoliton states, which are mechanically stable states with largest number of particles.
The presoliton state is formed by basic mechanically stable clusters composed of $\nb$ particles.
The knowledge of $\nb$ allows us to derive in Sec.~\ref{sec:minimal-overfilling}
the minimal overfilling necessary to generate solitons. In Sec.~\ref{sec:soliton-propagation} we describe how solitons propagate by periodic sequences of cluster movements and derive their time period and velocity. 
Thereafter we analyze in Sec.~\ref{sec:number-of-solitons} how many solitons form for a given overfilling 
and discuss in Sec.~\ref{sec:magic-sizes} how our results can be tested in experiments. We explain
the effective repulsive soliton-soliton interaction in Sec.~\ref{sec:soliton-interaction} and derive the particle currents mediated by solitons
in Sec.~\ref{sec:particle_currents}.

We use $U_0$, $\lambda$ and $\lambda^2/\mu U_0$ as units of 
energy, length and time in the following.
Our analysis is performed for hard-sphere diameters $0<\sigma<1$. Section~\ref{sec:sigma>1} discusses how results 
for $\sigma>1$ larger than the wavelength $\lambda=1$ are obtained from those for $0<\sigma<1$.

When referring to simulations, these are carried out by applying the recently developed method of 
Brownian cluster dynamics \cite{Antonov/etal:2022c}.

\section{Particle clusters in tilted periodic potential}
\label{sec:clusters}
An $n$-cluster is formed by $n$ particles in contact, i.e.\ with positions $x_1$, $x_1+\sigma,\ldots x_1+(n-1)\sigma$. We define
the position $x$ of the $n$-cluster as the position of the first particle, $x=x_1$. 
An $n$-cluster has size $n$ in terms of number of particles and it covers an interval of size $n\sigma$. 

If the particles in the $n$-cluster keep in contact, the mean force acting on it is 
$F_n(x)=f-\partial _x U_n(x)$, where
\begin{align}
U_n(x)&=\frac{1}{n} \sum\limits_{j=0}^{n-1} U(x+j\sigma)=
\frac{U_0}{2n}\sum\limits_{j=0}^{n-1} \cos[2\pi(x+j\sigma)]\nonumber\\
&=\frac{U_0\sin(\pi n \sigma)}{2 n \sin(\pi \sigma)}\cos[2\pi x + \pi(n\!-\!1)\sigma]
\label{eq:cluster-potential}
\end{align}
is the $n$-cluster potential. For the magic particle diameters $\sigma_n=(n-1)/n$,
$U_n(x)=0$, i.e.\ an $n$-cluster with particles of diameter $\sigma_n$ moves without surmounting barriers if the particles in the cluster stay together 
during the motion.

Particles in an $n$-cluster at position $x$ keep in contact if the non-splitting conditions
\begin{equation}
\frac{1}{l}\sum\limits_{j=0}^{l-1}F(x+j\sigma)>
\frac{1}{n\!-\!l}\sum\limits_{j=l}^{n-1}F(x+j\sigma)
\label{eq:nonsplitting-conditions}
\end{equation}
are obeyed, where $F(x)=F_1(x)=f-\partial_x U(x)$ is the force acting on a single particle; note that $f$ cancels
in these conditions.
The conditions mean the following: when
considering any decomposition of the $n$-cluster into an $l$-subcluster and $(n\!-\!l)$-subcluster at its left and right end, respectively, the velocity of (or force on) the $l$-subcluster must be larger than that of the $(n\!-\!l)$-subcluster, i.e.\
the subclusters do not separate.

If the non-splitting conditions are satisfied, the $n$-cluster at position $x$ has the velocity ($\mu=1$)
\begin{equation}
v_n(x)=F_n(x)=f-\frac{\partial U_n(x)}{\partial x}\,.
\end{equation}
If they are satisfied in an interval $[y,y']$ of positions, the time for the $n$-cluster to move from
$y$ to $y'$ is
\begin{equation}
\tau_n(y,y')=\int\limits_{y}^{y'}\frac{\dd x}{v_n(x)}\,.
\label{eq:taun}
\end{equation}

The tilted potential $U_n(x)-fx$ has barriers if the force $f$ is smaller than the
critical force 
\begin{equation}
\fc(\sigma,n)=\pi \frac{U_0}{n}\frac{|\sin(\pi n\sigma)|}{\sin(\pi\sigma)}
\label{eq:fc}
\end{equation}
for overtilting of the $n$-cluster potential. For a single-particle, $\fc(\sigma,1)=\pi U_0=\fc$, 
with $\fc$ from Eq.~\eqref{eq:fc-single}.
An $n$-cluster can be mechanically stable only for $f<\fc(\sigma,n)$.

It is mechanically stable if it is at a
position $x$, where $U_n(x)-fx$ has a local minimum, 
\begin{equation}
\partial_x U_n(x)=f\,,\quad \partial^2_x U_n(x)>0\,,
\label{eq:stable-mech-eq}
\end{equation}
and if the $n$-cluster does not fragment, i.e.\ if the 
nonsplitting conditions~\eqref{eq:nonsplitting-conditions} are obeyed.

Conditions~\eqref{eq:stable-mech-eq} are satisfied for positions $x=x_n^\pm(\sigma,f)+j$, where
\begin{align}
x_n^\pm(\sigma,f)=&
\,\frac{(1-n)\sigma}{2}+\frac{1}{2\pi}\arcsin\left(\frac{f}{\fc(\sigma,n)}\right)
\label{eq:x0pm-finite-f}\\[1ex]
&{}+\left\{\begin{array}{ll}
\tfrac{1}{2}\,, & \mbox{for $\sin(\pi n\sigma)>0$}\,, \\[2.5ex]
0\,, & \mbox{for $\sin(\pi n\sigma)$}<0\,.
\nonumber
\end{array}\right.
\end{align}
Those values $j$ are allowed for which $x_n^\pm(\sigma,f)+j\in[0,L[$. 
Inserting these values into the non-splitting conditions~\eqref{eq:nonsplitting-conditions}, one can decide whether
an $n$-cluster is mechanically stable.

\section{Presoliton states }
\label{sec:presoliton-states}
The presoliton state is the state of stable mechanical equilibrium with the largest number of particles. 
Simulations starting from different initial particle configurations show that it is formed by a sequence of 
evenly separated clusters with the same particle number $\nb$ and a residual of 
$n_{\rm res}<\nb$ particles, see Fig.~\ref{fig:illustration_open_problems}(a),
where $n_{\rm res}=1$; $n_{\rm res}=0$ is also possible. 
We call $\nb$ the basic stable cluster size and an $\nb$-cluster a basic one.
It will play a fundamental role in the following.

In Sec.~\ref{sec:presoliton-state-zero-force}, we first derive $\nb$ for the case of infinitesimal force $f=0^+$,
which we refer to as zero-force limit. 
In this limit, solitons can
occur only if the particle diameter $\sigma$ is a rational number 
(in units of the wavelength $\lambda=1)$,
\begin{equation}
\sigma_{p,q}=\frac{p}{q}\,,\quad p,q\in\mathbb{N}\,,\quad p<q\,.
\label{eq:sigmapq}
\end{equation}
This is a necessary condition, since the cluster potential $U_n(x)$ becomes zero
if  $n=q$ and $\sigma=p/q$. It means that
a $q$-cluster composed of particles of size $\sigma_{p,q}$ can move without surmounting barriers.
To represent each such $\sigma_{p,q}$ uniquely, $p$ and $q$ are taken to be 
coprime. Because $\sigma_{p,q}<1$, it is $q>p$.

The necessary condition $\sigma_{p,q}=p/q$
 does not imply that solitons must occur for all $p/q$.  
For example, in Ref.~\cite{Antonov/etal:2022a} solitons were studied for 
overfilling $\Delta=1$ and appeared for $p=q-1$ only.
We will show that solitons occur for most $p/q$, if the overfilling 
exceeds a minimal value $\Delta_{\rm min}$. This $\Delta_{\rm min}$
The required minimal overfilling is calculated in Sec.~\ref{sec:minimal-overfilling}.
It may, however, not be realizable for a finite system size $L$. 

The knowledge for $f=0^+$ allows us to determine $\nb$ for finite $f<\fc$ also, which we show
in Sec.~\ref{sec:presoliton-state-finite-force}.

\subsection{Basic stable cluster size $\nb$ for $f=0^+$}
\label{sec:presoliton-state-zero-force}
For an $n$-cluster to be in stable mechanical equilibrium, it must be
placed at a local minimum of the 
effective potential $U_n(x)$ [Eq.~\eqref{eq:cluster-potential}]  and it must not split [conditions~\eqref{eq:nonsplitting-conditions}].
We refer to these two conditions as translation stability and fragmentation stability. 
If both conditions are met, we call an $n$-cluster stable. 
An $n$-cluster is called stabilizable, if it is stable against fragmentation at a position
of a local minimum of $U_n(x)$.

A 1-cluster (single particle) is stabilizable, as it cannot fragment.
Clusters of size $n=q$ are unstable, because $U_q(x)$ has no local minimum. 
A cluster of size $n>q$ is unstable also. This is because
it can be divided into two subclusters, 
one to the left of size $(n\!-\!q)$, and one to the right of size 
$q$.  The right $q$-subcluster moves in the presence of an infinitesimal force $f$ 
and the left subcluster of size $(n\!-\!q)$ can only 
speed up the motion of the right $q$-subcluster. Hence, a cluster in 
stable mechanical equilibrium must have a size smaller than 
$q$. We conclude that $\nb\in\{1,\ldots,q\!-\!1\}$.

Interestingly, stabilizability of a cluster is related to a 
geometric property, which is the residual free space
when the cluster is accommodated in a minimal number of potential wells.
Specifically, we define the residual free space
$r_n$ of an $n$-cluster
as the difference between the space covered by $\lceil n\sigma_{p,q} \rceil$ accommodating potential wells 
($\lceil n\sigma_{p,q} \rceil$ wavelengths)
and the space $n\sigma_{p,q}$ covered by the cluster:
\begin{equation}
r_n=\lceil n\sigma_{p,q} \rceil-n\sigma_{p,q}\,.
\end{equation}
Here, $\lceil x\rceil$ is the smallest integer larger than $x$.

The relation between stabilizability and residual free space 
is given by the following free-space theorem, 
derived in \ref{app:proof-free-space-theorem}:
\textit{If an isolated $n$-cluster is stabilizable, its residual free space $r_n$ is smaller than that of any
cluster of smaller size, i.e.\ it holds}
\begin{equation}
r_n<r_l\,,\hspace{1em}l=1,\ldots,n-1\,.
\label{eq:free-space-theorem}
\end{equation}
It follows that the largest among the stabilizable clusters has smallest residual free space.
This is because for any two stabilizable $n$- and $n'$-clusters with sizes $n>n'$, it holds $r_n<r_{n'}$ according to Eq.~\eqref{eq:free-space-theorem}. 

The cluster with minimal residual free space is unique, i.e.\ the $r_n$ are all different for $n\in\{1,\ldots,q-1\}$.
To show this, let us assume that there exist $n_1,n_2\in\{1,\ldots,q\!-\!1\}$ with 
$n_2<n_1$ and
$r_{n_1}=\lceil n_1\sigma_{p,q} \rceil-n_1\sigma_{p,q}=\lceil n_2\sigma_{p,q} \rceil-n_2\sigma_{p,q}=r_{n_2}$, i.e.\
$(n_1-n_2)\sigma_{p,q}=\lceil n_1\sigma_{p,q} \rceil-\lceil n_2\sigma_{p,q} \rceil$.
This would imply that $(n_1\!-\!n_2)\sigma_{p,q}=(n_1\!-\!n_2)p/q$ is an integer. 
However, this is impossible for coprime $p$ and $q$, and
$(n_1\!-\!n_2)\in\{1,\ldots,q\!-\!2\}$. In particular, we obtain
\begin{equation}
\min\{r_1,\ldots r_{q-1}\}=\frac{1}{q}\,.
\label{eq:min-r}
\end{equation}

A mechanically stable state with largest particle number has highest coverage 
$N\sigma_{p,q}/L$. The highest coverage is obtained, if successive potential wells accommodate
a largest stabilizable cluster with minimal residual free space.
Accordingly, 
$\nb$ is determined from a principle of minimum residual free space:
\begin{equation}
\nb(\sigma_{p,q})=\hspace{-0.3em}\argmin_{n\in\{1,\ldots,q-1\}}\hspace{-0.1em}(\lceil n\sigma_{p,q} \rceil-n\sigma_{p,q})\,,
\label{eq:nast-determ}
\end{equation}
where $\argmin(f(x))$ gives the argument $x$ at which the function $f(x)$ has its minimum.
Hence, $\nb(\sigma_{p,q})$
is equal to that $n\in\{1,\ldots,q-1\}$, where $f(n)=\lceil n\sigma_{p,q} \rceil-n\sigma_{p,q}$ is minimal.
Equivalently, $\nb$ is determined by Eq.~\eqref{eq:min-r}, i.e.\
$r_{\nb}=\lceil \nb\sigma_{p,q} \rceil-\nb\sigma_{p,q}=1/q$. This condition can be rewritten in the form
\begin{equation}
q\hspace{0.1em}l-p\hspace{0.1em}\nb=1\,,
\label{eq:diophantine}
\end{equation}
where $l=\lceil \nb p/q\rceil$. 

For $l$ being an arbitrary integer, Eq.~\eqref{eq:diophantine} is a linear Diophantine equation in
the two variables $\nb$ and $l$. Its solutions are \cite{Vorobyov:1980}
\begin{align}
\nb&=-p^{\varphi(q)-1}+qj\,,
\label{eq:nast-sol}\\
&l=-\frac{p^{\varphi(q)}-1}{q}+pj\,,
\label{eq:l-sol}
\end{align}
where $j$ can be any integer $j\in\mathbb{Z}$, 
$\varphi(.)$ is Euler's Phi function \cite{Abramowitz/Stegun:1965} and $(p^{\varphi(q)}-1)/q$ is an integer due to 
the Euler-Fermat theorem \cite{Vorobyov:1980}. Because $\nb\in\{1,\ldots,q-1\}$, the integer $j$ in Eq.~\eqref{eq:nast-sol}
must be $j=\lceil p^{\varphi(q)-1}/q\rceil$. We hence obtain the explicit solution
\begin{equation}
\nb(\sigma_{p,q})=q\left\lceil \frac{p^{\varphi(q)-1}}{q} \right\rceil-p^{\varphi(q)-1}\,.
\label{eq:nb-solution}
\end{equation}
The $j=\lceil p^{\varphi(q)-1}/q\rceil$ gives also the required $l=\lceil \nb p/q\rceil$:
\begin{align}
\left\lceil \nb\frac{p}{q}\right\rceil&=\left\lceil \left\lceil\frac{p^{\varphi(q)-1}}{q}\right\rceil p-\frac{p^{\varphi(q)}}{q}\right\rceil
\nonumber\\
&=\left\lceil \left\lceil\frac{p^{\varphi(q)-1}}{q}\right\rceil p-\frac{p^{\varphi(q)}-1}{q}-\frac{1}{q}\right\rceil
\nonumber\\
&=-\frac{p^{\varphi(q)}-1}{q}+p\left\lceil\frac{p^{\varphi(q)-1}}{q}\right\rceil=l\,.
\label{eq:lcheck}
\end{align}
Here we have used $\lceil x+j\rceil=\lceil x\rceil+j$ for integer $j$ and $\lceil -1/q\rceil=0$.
The last equality in Eq.~\eqref{eq:lcheck} follows when inserting $j=\lceil p^{\varphi(q)-1}/q\rceil$ into Eq.~\eqref{eq:l-sol}.
Let us note also that Eq.~\eqref{eq:diophantine} implies that $\nb$ and $q$ must be coprime.

%%%%%%%%%%%%%%%%%%%%%%%%%%%%%%%%%%%%%%%%%%%%%%%
\begin{figure}[t!]
\centering
\includegraphics[width=\columnwidth]{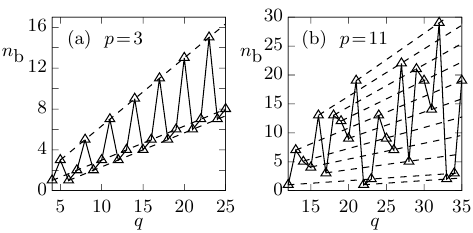}
\caption{Size $\nb$ of basic stable clusters [Eq.~\eqref{eq:nb-solution}] 
for different rational particle diameters $\sigma_{p,q}=p/q$, when $p$ is fixed and $q$ is varied. In (a) $p=3$ and in (b) $p=11$.
In the graphs, results are shown also for non-coprime $p$ and $q$. In that cases, $p'=p/\gcd(p,q)$ and $q'=q/\gcd(p,q)$ 
have to be used in Eq.~\eqref{eq:nb-solution}, where $\gcd(p,q)$ is the greatest common divisor of $p$ and $q$.
Dashed lines connect symbols for $q$-values differing by integer multiples of $p$.}
\label{fig:nstar}
\end{figure}
%%%%%%%%%%%%%%%%%%%%%%%%%%%%%%%%%%%%%%%%%%%%%%%

Figure~\ref{fig:nstar} shows representative results of $\nb$ for $\sigma_{p,q}=p/q$ in dependence of $q$ for (a) $p=3$ and
(b) $p=11$. They exhibit a recurrent behavior with a period $p$, where $\nb(\sigma_{p,(q+jp)})$
increases linearly with $j$ with a slope that is an integer multiple of $1/p$, see the dashed lines. 

We note that $\nb$ can be equal to one, which means that all $n$-clusters with $n\ge2$ are mechanically unstable.
A particular set of particle diameters with $\nb(\sigma_{p,q})=1$ is given by $\sigma_{p,q}=(q-1)/q$.
For $\sigma_{p,q}=1/q$, $\nb=q-1$.

\subsection{Basic stable cluster size $\nb$ for $f>0$}
\label{sec:presoliton-state-finite-force}
For drag forces $f>0$, the largest translational stable cluster has size
\begin{equation}
n_{\rm max}(\sigma,f)=\max\{n\in\mathbb{N}\,|\, f< \fc(\sigma,n)\}\,,
\label{eq:nmax}
\end{equation}
where the critical force for overtilting is given in Eq.~\eqref{eq:fc}.
This limits the range $1,\dots,n_{\rm max}(\sigma,f)$ of
stabilizable clusters.

Interestingly, our simulations show that Eq.~\eqref{eq:free-space-theorem} remains valid for $f>0$, with $\sigma$ replacing 
$\sigma_{p,q}$. Hence the residual free space of a stabilizable cluster with size $n$, $n\le n_{\rm max}(\sigma,f)$,
is smaller than that of clusters of size smaller than $n$. Similarly as in Eq.~\eqref{eq:nast-determ},
\begin{equation}
\nb(\sigma,f)=\hspace{-0.3em}
\argmin_{n\in\{1,\ldots,n'_{\rm max}\}}\hspace{-0.2em}(\lceil n\sigma\rceil-n\sigma)\,.
\label{eq:nast-determ2}
\end{equation}
Here, $n'_{\rm max}\le n_{\rm max}(\sigma,f)$ is the size of the largest stabilizable cluster, i.e.\ which at a translational stable position according 
to \eqref{eq:stable-mech-eq} fulfills the non-splitting conditions \eqref{eq:nonsplitting-conditions}.

Equations~\eqref{eq:nmax} and \eqref{eq:nast-determ2} 
imply that $\nb(\sigma,f)$ can be determined for any 
$\sigma<1$ and $0<f<\fc$ by the following method: First one checks whether the $n_{\rm max}$-cluster, with $n_{\rm max}(\sigma,f)$
from Eq.~\eqref{eq:nmax}, is stabilizable, i.e.\ whether it satisfies the non-splitting 
conditions \eqref{eq:nonsplitting-conditions}. If it is not stabilizable, $n$ is decreased
by one and the stabilizability of this $n$-cluster is checked. The procedure is repeated until the cluster of size $n$
is stabilizable. This $n$ is equal to $\nb$,
since by decreasing the cluster size, the residual free space increases.

%%%%%%%%%%%%%%%%%%%%%%%%%%%%%%%%%%%%%%%%%%%%%%%
\begin{figure*}[t!]
\centering
\includegraphics[width=\textwidth]{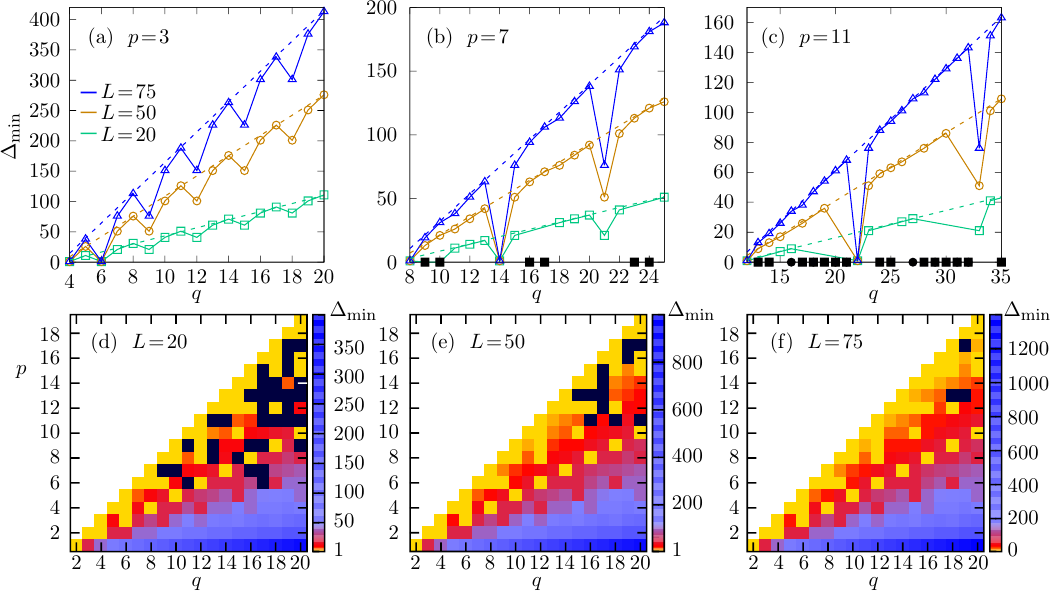}
\caption{Minimal overfilling $\Delta_{\rm min}(\sigma_{p,q},L)$ for soliton appearance [Eq.~\eqref{eq:Deltamin-determ}]
for different rational particle diameters $\sigma_{p,q}=p/q$ and different system sizes $L$. 
In (a)-(c), $\Delta_{\rm min}(\sigma_{p,q},L)$ is shown in dependence of $q$ for fixed (a) $p=3$, (b) $p=7$, and (c)
$p=11$, and $L=20$, 50, and 75. 
Dashed lines with slopes $L/p$ show the overall linear increase of $\Delta_{\rm min}$ with $q$.
Black symbols mark $q$ values where no solitons occur (squares for L=20, circels for $L=50$).
In (d)-(f), $\Delta_{\rm min}(\sigma_{p,q},L)$ is shown for different $p$ and $q$ in a color-coded representation for (d) $L=20$, 
(e) $L=50$, and (f) $L=75$. Black squares represent $(q,p)$-values, where the self-consistency 
condition~\eqref{eq:Deltamin-selfconsistency} is violated and no solitons form.}
\label{fig:minimal_overfilling_zero_f}
\end{figure*}
%%%%%%%%%%%%%%%%%%%%%%%%%%%%%%%%%%%%%%%%%%%%%%%

\section{Minimal overfilling}
\label{sec:minimal-overfilling}
Knowing $\nb$, the minimal overfilling
$\Delta_{\rm min}=\Delta_{\rm min}(\sigma,L)$ for soliton appearance
follows from geometric considerations. In the presoliton state,
the maximal number $N_{\rm pre}^{\rm b}$ of stable $\nb$-clusters fitting into a system of length $L$ is 
\begin{equation}
N_{\rm pre}^{\rm b}=\left\lfloor \frac{L}{\mb}\right\rfloor\,,
\label{eq:Nbpre}
\end{equation}
where $\lfloor x\rfloor$ is the largest integer smaller than $x$.
The number of potential wells accommodating all 
$\nb$-clusters is 
\begin{equation}
M_{\rm pre}=N_{\rm pre}^{\rm b} \mb\,.
\label{eq:Mpre}
\end{equation}
There can be residual potential wells not accommodating $\nb$-clusters. 
Their number is 
\begin{equation}
m_{\rm res}=L-M_{\rm pre}=L\ourmod \mb\,,
\label{eq:Mrest}
\end{equation}
where $a \ourmod b=a-\lfloor a/b\rfloor b$ denotes the modulo operation.
The number of particles fitting into the $m_{\rm res}$ residual wells is 
\begin{equation}
n_{\rm res}=\left\lfloor \frac{m_{\rm res}}{\sigma}\right\rfloor
=\left\lfloor \frac{L\ourmod \mb}{\sigma}\right\rfloor\,.
\label{eq:Nrest}
\end{equation}
Accordingly, the maximal particle number in a configuration 
of stable mechanical equilibrium is 
\begin{equation}
N_{\rm pre}=N_{\rm pre}^{\rm b}\,\nb+n_{\rm res}\,.
\label{eq:nast}
\end{equation}
Adding one more particle gives rise to a soliton, i.e.\ 
\begin{align}
\Delta_{\rm min}(\sigma,L)&=N_{\rm pre}+1-L
\label{eq:Deltamin-determ}\\
&=\left\lfloor \frac{L}{\mb}\right\rfloor \nb
+\left\lfloor \frac{L\ourmod \mb}{\sigma}\right\rfloor+1-L\,.\nonumber
\end{align}
However, the coverage $[\Delta_{\rm min}(\sigma,L)+L]\sigma$ by the particles must not exceed the system length $L$, 
\begin{equation}
[\Delta_{\rm min}(\sigma,L)+L]\sigma< L\,.
\label{eq:Deltamin-selfconsistency}
\end{equation}
If this self-consistency condition is not fulfilled, solitons do not form.

Let us discuss a few examples for the case of weak forces, where 
solitons can only occur when $\sigma$ is close to $\sigma_{p,q}=p/q$.
We therefore consider the zero-force limit $f=0^+$.

%%%%%%%%%%%%%%%%%%%%%%%%%%%%%%%%%%%%%%%%%%%%%%%
\begin{figure*}[t!]
\centering
\includegraphics[width=0.85\textwidth]{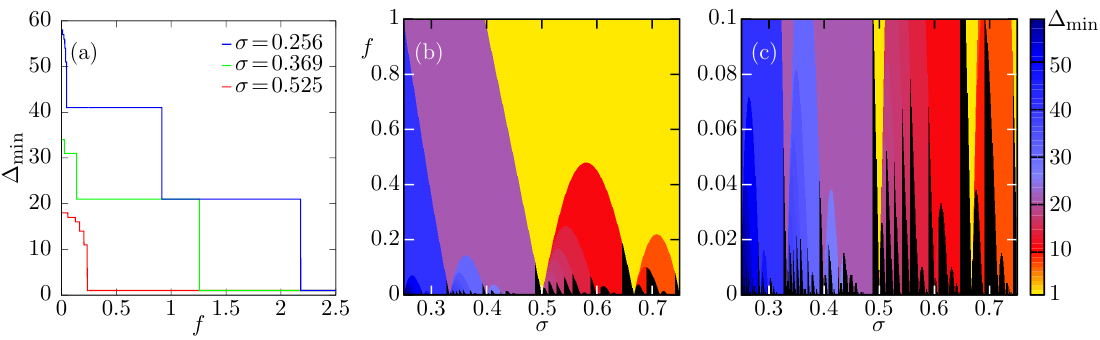}
\caption{Minimal overfilling $\Delta_{\rm min}$
required for solitons to form as a function of drag force $f$ and particle diameter $\sigma$ in a system of length $L=20$.
In (a) the stepwise decrease of $\Delta_{\rm min}$ with $f$ is demonstrated
for three particle diameters $\sigma$. Panel (c) is a zoom-in of the small-$f$ regime ($f\le0.1$) of panel (b), showing
a complex fractal-like pattern. Results were obtained by numerically solving Eqs.~\eqref{eq:langevin} in the zero-noise limit.
In the black areas, solitons do not form.}
\label{fig:minimal_overfilling_finite_f}
\end{figure*}
%%%%%%%%%%%%%%%%%%%%%%%%%%%%%%%%%%%%%%%%%%%%%%%

For a particle diameter $\sigma_{p,q}\!=\!3/5$ and a system size $L\!=\!20$,
$\lceil n\sigma_{p,q} \rceil-n\sigma_{p,q}=2/5,4/5,1/5$ and 3/5 for $n=1,2,3$, and 4. Hence
$\nb=3$,
i.e.\ the minimum free space in accommodating potential wells is obtained for cluster size three. 
This is in agreement with Eq.~\eqref{eq:nb-solution}.
Hence, $N_{\rm pre}^{\rm b}=\lfloor 20/\lceil 9/5\rceil\rfloor=20/2=10$ from Eq.~\eqref{eq:Nbpre}, 
$M_{\rm pre}=10\lceil 9/5\rceil=20$ from Eq.~\eqref{eq:Mpre}, $m_{\rm res}=n_{\rm res}=0$ from Eqs.~\eqref{eq:Mrest}, \eqref{eq:Nrest},
$\nb=30$ from Eq.~\eqref{eq:nast}, yielding $\Delta_{\rm min}=11$ according to Eq.~\eqref{eq:Deltamin-determ}. The self-consistency 
condition~\eqref{eq:Deltamin-selfconsistency} reads $(11+20)(3/5)=93/5<20$ and is satisfied. 

The particle size $\sigma_{p,q}=3/11$  and system size 
$L=23$ give an example, where not all particles are part of $\nb$-clusters in the presoliton state. 
Here $\nb=7$, $n_{\rm res}=3$, and $\Delta_{\rm min}=58$, which satisfies the 
self-consistency condition~\eqref{eq:Deltamin-selfconsistency}.

An example, where soliton formation is impossible in a system of size $L=20$ is, when the particles have 
diameter $\sigma_{p,q}=7/9$. In that case, $\nb=5$ and $\Delta_{\rm min}=6$, which violates the self-consistency 
condition~\eqref{eq:Deltamin-selfconsistency},
$[\Delta_{\rm min}(\sigma_{p,q},L)+L]\sigma_{p,q}=(6+20)7/9>20$.

Figure~\ref{fig:minimal_overfilling_zero_f} shows the minimal overfilling $\Delta_{\rm min}(\sigma_{p,q},L)$ for soliton appearance 
for different $\sigma_{p,q}=p/q$ and $L$. 
In simulations we have verified both the absence of soliton formation and
the values predicted by Eq.~\eqref{eq:Deltamin-determ}.

%%%%%%%%%%%%%%%%%%%%%%%%%%%%%%%%%%%%%%%%%%%%%%%
\begin{figure*}[t!]
\centering
\includegraphics[width=0.9\textwidth]{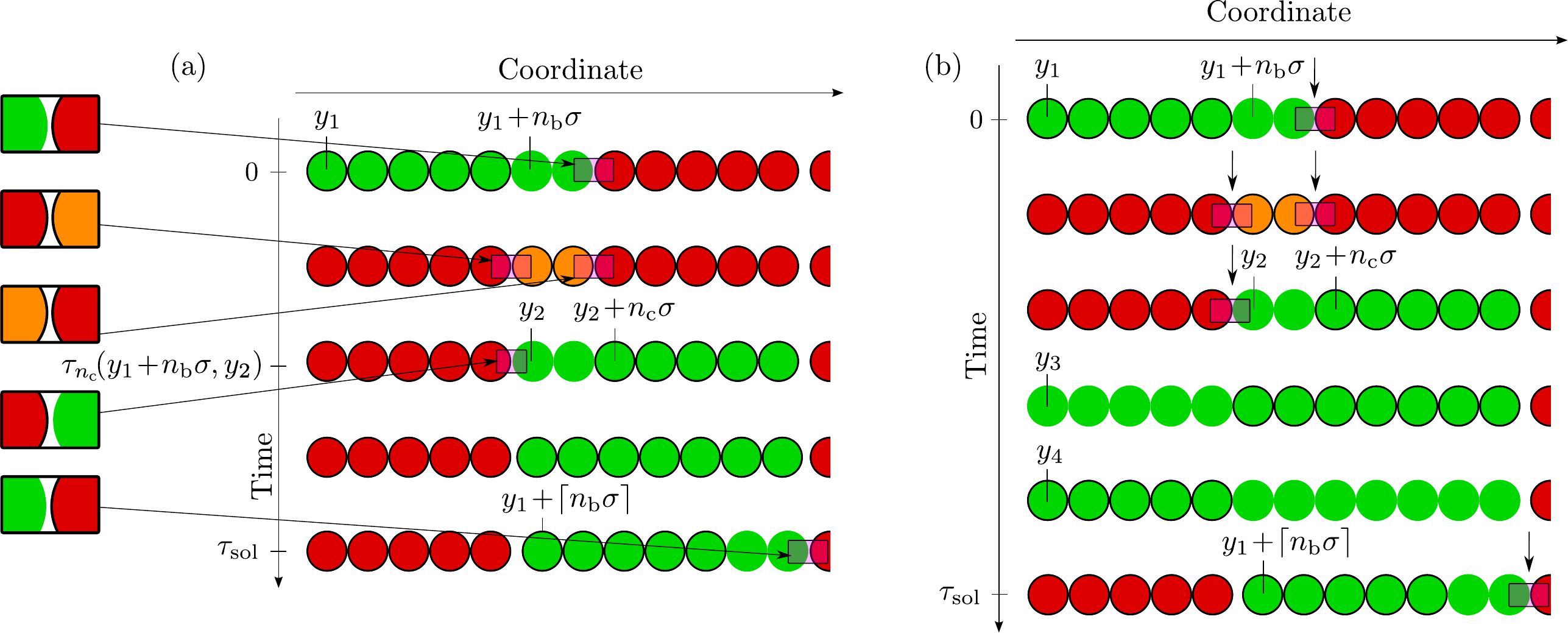}
\caption{Soliton propagation in zero-noise limit at drag force $f=0.01$ for particle diameters (a) $\sigma=0.57$ and (b) $\sigma=0.5725$.
For both $\sigma$, sizes of the basic stable and core soliton cluster are $\nb=5$ and $\nc=2$.
$\nb$-clusters are marked in red and relax toward positions of stable mechanical equilibria. 
The soliton consists of a periodic sequence
of cluster movements. 
The core $\nc$-cluster (orange) is the smallest in the sequence, and the larger soliton clusters (green) are resulting 
from mergers of the core cluster with $\nb$-clusters. Cluster sizes in the sequence change due to attachment and 
detachment processes, where particles of detaching/attaching clusters are depicted without black circle borders.
For the particle diameter in (a), propagation mode A occurs: at time $t=0$, a composite $(\nb+\nc)$-cluster 
reaches position $y_1$, where a core $\nc$-cluster detaches from it at $y_1+\nb\sigma$. When the core cluster reaches 
$y_2$ at time $\tau_{\nc}(y_1+\nb\sigma,y_2)$, it attaches to an
$\nb$-cluster at $y_2+\nc\sigma$. Thereafter the composite $(\nc+\nb)$-cluster formed by the attachment moves until
it reaches $y_1+\mb$ at time $\tau_{\rm sol}=\tau_{\nc}(y_1+\nb\sigma,y_2)+\tau_{\nc+\nb}(y_2,y_1+\mb)$. 
At this point, one period of the soliton motion is completed: the soliton state is equivalent to that at
time $t=0$, where a core $\nc$-cluster detaches from the composite $(\nc+\nb)$-cluster at its right end.
For the slightly larger particle diameter in (b), the propagation mode B occurs: before the composite 
$(\nc+\nb)$-cluster reaches $y_1+\mb$, the relaxing $\nb$-cluster to its left attaches and detaches 
from it. Spacings between soliton clusters and $\nb$-clusters in this example are very small. Enlargements of the parts 
marked by rectangles in (a) point to the cluster separation. In (b), gaps between clusters in the rectangles
are indicated by vertical arrows.}
\label{fig:illustration-soliton-propagation-modes}
\end{figure*}
%%%%%%%%%%%%%%%%%%%%%%%%%%%%%%%%%%%%%%%%%%%%%%%

Figures~\ref{fig:minimal_overfilling_zero_f}(a)-(c) show the dependence of $\Delta_{\rm min}(\sigma_{p,q},L)$ 
when varying $q$ at fixed $p=3$, 7, and 11 for system sizes $L=20$, 50 and 75. As a function of $q$,
$\Delta_{\rm min}$ shows an alternating behavior of increase and decrease with an overall linear increase, 
where the slope of the linear
dependence rises with $L$. Equation~\eqref{eq:Deltamin-determ} predicts an overall behavior $\Delta_{\rm min}\sim (1/\sigma-1)L=(q/p-1)$, i.e.\ the 
slope should be $L/p$. The dashed lines in Fig.~\ref{fig:minimal_overfilling_zero_f}(a)-(c) representing 
$(q/p-1)L$ indeed capture the overall linear increase of $\Delta_{\rm min}$ with $q$. 

Remarkable are the particle diameters $\sigma_{p,q}=p/q$ for $p=7$ and  $p=11$
where no solitons form 
for small drag force $f$. 
At these particle diameters, 
there is not enough free space left to generate a soliton by adding one particle
to the presoliton state. These states of soliton absence are represented 
by black squares in Figs.~\ref{fig:minimal_overfilling_zero_f}(e)-(f), where  $\Delta_{\rm min}(\sigma_{p,q},L)$
is plotted in a color-coded representation for a section of the $(p,q)$-grid 
and three different system sizes $L=20$ [Fig.~\ref{fig:minimal_overfilling_zero_f}(e)],
$L=50$  [Fig.~\ref{fig:minimal_overfilling_zero_f}(d)], and $L=75$ [Fig.~\ref{fig:minimal_overfilling_zero_f}(f)].
Theoretical results in Fig.~\ref{fig:minimal_overfilling_zero_f} were checked against simulations.

To exemplify the behavior of $\Delta_{\rm min}(\sigma,f)$ for a wide range 
of particle diameters and drag forces, we have carried out extensive simulations for a system of size $L=20$. The results are
displayed in Fig.~\ref{fig:minimal_overfilling_finite_f}(a)-(c), where
in the color-coded representation of Fig.~\ref{fig:minimal_overfilling_finite_f}(b)  a resolution
$\Delta f=10^{-3}$ and $\Delta\sigma=10^{-3}$ was chosen. Figure~\ref{fig:minimal_overfilling_finite_f}(c)
depicts the region with small $f$-values enlarged. 
For hundred randomly chosen points in the  $\sigma$-$f$-plane
in Fig.~\ref{fig:minimal_overfilling_finite_f}(b), we in particular 
checked that the simulated values agree with the 
predicted ones according to the algorithmic procedure described after Eq.~\eqref{eq:nast-determ2}.

Figure~\ref{fig:minimal_overfilling_finite_f}(a) shows how
$\Delta_{\rm min}(\sigma,f)$ changes with $f$ for three fixed $\sigma$ and
$L=20$. The overfilling at small $f$ is larger for smaller $\sigma$.
With increasing $f$, $\Delta_{\rm min}(\sigma,f)$ decreases in a stepwise manner, with the steps occurring at different
$f$ for different $\sigma$. For large $f$, $\Delta_{\rm min}$ becomes one for all $\sigma$.

A complete picture of soliton formation in the range $0.25\le\sigma\le 0.75$ and $0<f\le1$ is 
given in Fig.~\ref{fig:minimal_overfilling_finite_f}(b). It has a remarkable complex structure, where black 
color marks soliton absence. At fixed $f$, $\Delta_{\rm min}(\sigma,f)$ can change often between small and large
values when $\sigma$ is varied. The black areas occur at small $f$, meaning
that solitons formation requires a minimal $f$ for non-magic $\sigma$.
Figure~\ref{fig:minimal_overfilling_finite_f}(c) is a zoom-in of the region $f\le0.1$, revealing
a fractal-like pattern. When increasing $L$, the black area becomes smaller and it vanishes for $L\to\infty$.

\section{Soliton propagation}
\label{sec:soliton-propagation}

\subsection{Propagation modes}
\label{subsec:soliton-modes}
A soliton propagates 
by a sequential movement of clusters formed by splitting (detachment) and merging (attachment) events. 
There are two propagation modes, the basic one A and a variant B. Whether one or the other mode occurs, depends on
$\sigma$ and $f$. The variant of the basic mode is less frequently encountered, in particular for $f\ll\fc$. 
Both modes can be described by two clusters, the core soliton cluster of size $\nc$,
and the composite soliton cluster of size $\nc+\nb$.
Among the clusters involved in the soliton propagation, the
core cluster has smallest size.

Figure~\ref{fig:illustration-soliton-propagation-modes}(a) illustrates the basic mode A: when a composite 
$(\nc+\nb)$-cluster terminates its movement at a position $y_1$,
an $\nc$-cluster detaches at its right end, leaving an $\nb$-cluster behind.
After the detachment, the $\nb$-cluster starts to relax towards a position of stable mechanical equilibrium, 
and the $\nc$-cluster moves until
reaching a position $y_2$, where it attaches to an $\nb$-cluster. The composite $(\nc+\nb)$-cluster formed by the attachment
moves until reaching a position $y_1+\mb$. This completes one period of the soliton propagation: at the position 
$y_1+\mb$ equivalent to $y_1$, the $(\nc+\nb)$-cluster terminates its movement, 
because an $\nb$-cluster detaches at its right end. Accordingly, a soliton moves a distance $\mb$ in one period.

In the variant B of the basic propagation mode, the relaxing $\nb$-cluster is attaching to and shortly after detaching from the 
composite cluster, see Fig.~\ref{fig:illustration-soliton-propagation-modes}(b).
This slight modification of the basic propagation mode only occurs if at the time instant of the composite cluster formation,
the distance between the right end of the relaxing $\nb$-cluster and the left end of the composite cluster is very small and
the relaxing cluster moves faster than the composite cluster. In the short intermediate time interval
between detachment and attachment of the $\nb$-cluster, the soliton propagation is mediated by an $(\nb+\nc+\nb)$-cluster.

Movie~2 in the SM \cite{suppl-solitary-cluster-waves} shows the soliton propagation for modes A and B
described in Figs.~\ref{fig:illustration-soliton-propagation-modes}(a) and (b).
The small change of $\sigma$ from 0.57 to 0.5725 changes the soliton mode but not the cluster sizes $\nb$ and $\nc$.
In such cases, the soliton velocity is significantly larger for the B mode, as can be seen also in the movie.

In what follows, we focus on the basic propagation mode A. 
Quantities like soliton potentials or velocities discussed below can
be treated analogously for the variant B. For the
velocity of type B solitons, we give the corresponding calculation in \ref{app:vsol-type2-solitons}.

For weak drag force $f$, or, strictly speaking in the zero-force limit, the composite cluster has size 
$\nc+\nb=q$, i.e.\ it is the cluster that can move without surmounting barriers. 
This in particular implies that $\nc+\nb$ and $\nb$ are coprime, see comment after Eq.~\eqref{eq:lcheck}.
The core soliton cluster then has size
\begin{equation}
\nc=q-\nb\hspace{1em}\mbox{for}\hspace{0.3em}f=0^+\,.
\label{eq:nc-zero-force-limit}
\end{equation}
The motion of the $q$-cluster, however, does not span a full wavelength of the potential. 
This is because it splits into an $\nb$- and 
$\nc$-cluster at some point, which is $y_1$ in Fig.~\ref{fig:illustration-soliton-propagation-modes}(a). 
At this point, the non-splitting conditions~\eqref{eq:nonsplitting-conditions} are violated.

For larger $f>0$, the core cluster can have a size smaller than $q-\nb$.
Other than in the $f=0^+$ case, where $U_q(x)$ is constant for $\sigma=\sigma_{p,q}$,
it is possible that neither the core nor the composite cluster are
able to move over one period, even if the non-splitting conditions were obeyed everywhere. 
This is demonstrated in Fig.~\ref{fig:soliton-potential}(a), where we show the two tilted cluster potentials
$U_{\nc}(x)-fx$ and $U_{\nc+\nb}(x)-fx$ of the core and composite cluster for 
$\sigma = 0.57$ and $f=0.01$.
Both tilted cluster potentials exhibit barriers, i.e.\ neither of the 
two soliton clusters could move over a full wavelength. 

Due to the phase shift of the two cluster potentials, the barriers occur at different points.
This enables the two soliton clusters to move over consecutive intervals of the period, where the forces 
$f-U'_{\nc}(x)$ and $f-U'_{\nc+\nb}(x)$ 
in each interval are positive and the non-splitting conditions~\eqref{eq:nonsplitting-conditions} 
are fulfilled. One may view this as in a relay race, where the relay is passed between $\nc$ and $(\nc+\nb)$-clusters.

As $\nb$ is known, it is possible to calculate $\nc$
based on the tilted cluster potentials: one needs to
determine  that $n$, where the total force acting on an $n$-cluster in one part of the period and on an $(n+\nb)$-cluster 
in the other part of the period is positive and 
where the non-splitting conditions~\eqref{eq:nonsplitting-conditions} in both parts are satisfied. 
This $n$ is equal to $\nc$. For $\nb=1$, this gives  \cite{Antonov:2023}
\begin{align}
\hspace*{-0.75em}\nc =&\min
\left\{n\in\mathbb{N}\,\hspace{-1.5em}\left.\phantom{\frac{1}{2\pi}}\right|\,\left[\frac{1}{2\pi}\hspace{-0.1em}\arcsin(f/\fc)+n\right]>
\left[\phantom{\frac{1}{2\pi}}\hspace{-1.4em}(n\!+\!1)\sigma
\right.\right.\hspace*{0.5em}\\
&\hspace{-2.25em}\left.\left.
+\frac{1}{2\pi}\arccot\left(\frac{\pi \sin(\pi \sigma)}{\sin(\pi n \sigma)\sin[\pi (n\!+\!1) \sigma]}
\!-\! \cot[\pi(n\!+\!1)\sigma]\right)\right]\!\right\}%,
\nonumber
\end{align}
with $\arccot(x)\hspace{-0.3em}\in\,]-\pi/2,0[$ for $x\!<\!0$ and $\arccot(x)\hspace{-0.3em}\in\,]0,\pi/2[$ for $x\ge0$.

We show below, see Eq.~\eqref{eq:deltaNcl-deltaNsol-det2}, 
that $\nc$ has a value giving coprime $\nb$ and $(\nb+\nc)$. 

\subsection{Soliton potential and force field}
\label{subsec:soliton-fields}
\begin{figure}[t!]
\includegraphics[width=\columnwidth]{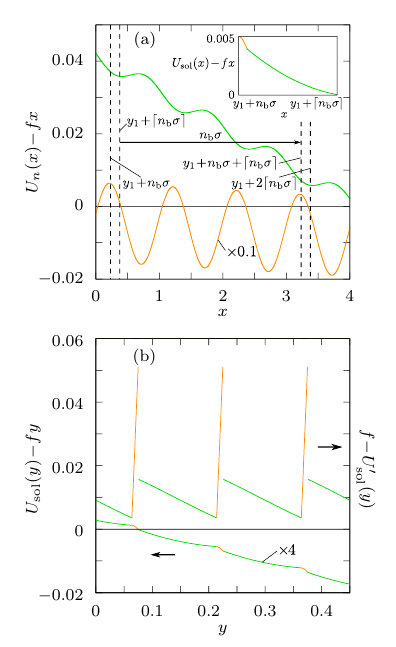}
\caption{(a) Tilted potentials $U_{\nc+\nb}(x)-fx$ (green) and $U_{\nc}-fx$ (orange) 
of the composite $(\nc+\nb)$- and core $\nc$-cluster
for $\sigma = 0.57$ and $f = 0.01$, where $\nb= 5$ and $\nc= 2$. The clusters of sizes $\nb$ and $(\nc+\nb)$
have positions in the intervals bounded by the vertical dashed lines. When an $(\nc+\nb)$-cluster reaches position
$y_1+\mb$, an $\nc$-cluster detaches at its right end and the soliton propagation continues with this
$\nc$-cluster at position $y_1+\mb+\nb\sigma$. The corresponding jump in soliton position of size $\nb\sigma$ is indicated by the
horizontal arrow. The inset shows the soliton potential \eqref{eq:Usol} in the interval
$[y_1+\nb\sigma, y_1+\mb[$.
(b) Tilted soliton potential $U_{\rm sol}(y)-fy$ and corresponding force field in the continuous-coordinate representation.
For this representation, the potential 
$U_{\rm sol}(y)$ from the interval $[y_1+\nb\sigma, y_1+\mb[$ in (a) is periodically continued.
Parts stemming from the composite and core soliton cluster are 
marked in green and orange, respectively.}
\label{fig:soliton-potential}
\end{figure}

Taking the soliton position as that of the soliton clusters, i.e.\ the position $x$ of the leftmost particle of the 
$\nc$- and $(\nc+\nb)$-clusters in the basic propagation mode, we can define a potential for the
soliton motion for $y\in[y_1+\nb\sigma, y_1+\mb[$:
\begin{linenomath*}
\begin{align}
U_{\rm sol}(y)=&\left[U_{\nc}(y)-U_{\nc}(y_2)\right]\Theta(y_2-y)\nonumber\\
&{}+\left[U_{\nc+\nb}(y) - U_{\nc+\nb}(y_2)\right]\Theta(y-y_2)\,.
\label{eq:Usol}
\end{align}
\end{linenomath*}
Here, $\Theta(.)$ is the Heaviside step function [$\Theta(y)=1$ for $y\ge1$ and zero otherwise].
The constants $U_{\nc}(y_2)$ and $U_{\nc+\nb}(y_2)$ were subtracted from the cluster potentials
to make the soliton potential continuous at $y=y_2$.
That way, the force $f-U'_{\rm sol}(y)$ on the soliton jumps at $y=y_2$
from $f-U'_{\nc}(y_2)$ to
$f-U'_{\nc+\nb}(y_2)$, but has no $\delta$-function singularity.
The tilted soliton potential  $U_{\rm sol}(y)-fy$
is shown in the inset of Fig.~\ref{fig:soliton-potential}(a),
where the parts stemming from the core and composite soliton clusters are marked 
as in the main figure.

When a soliton starting at $y_1+\nb\sigma$ has moved one time period $\tau_{\rm sol}$, its position jumps from 
$(y_1+\mb)$ to $y_1+\mb+\nb\sigma$, because when the $(\nc+\nb)$-cluster reaches position $(y_1+\mb)$,
an $n_c$-cluster at its right end detaches, which then becomes the soliton cluster at position $y_1+\mb+\nb\sigma$, see
Fig.~\ref{fig:soliton-potential}(a).
After the jump of size $\nb\sigma$, the potential $U_{\rm sol}$ is the same
as at the starting position.

%%%%%%%%%%%%%%%%%%%%%%%%%%%%%%%%%%%%%%%%%%%%%%%
%%%%%%%%%%%%%%%%%%%%%%%%%%%%%%%%%%%%%%%%%%%%%%%

For displaying the soliton potential,
it is helpful to use a continuous-coordinate representation where the jumps in the real soliton coordinate are removed. 
This means that we periodically continue $U_{\rm sol}(y)$ from the interval $y\in[y_1+\nb\sigma, y_1+\mb[$ in
Eq.~\eqref{eq:Usol}.
The soliton potential in the continuous coordinate representation hence is periodic with a period length 
equal to the residual free space  $r_{\nb}=\mb-\nb\sigma$ of the basic stable cluster.

Figure~\ref{fig:soliton-potential}(b) shows the tilted soliton potential $U_{\rm sol}(y)-fy$ 
for three periods $r_{\nb}$ of $U_{\rm sol}(y)$,
in the continuous-coordinate representation, 
and the corresponding force field $f-U'_{\rm sol}(y)$.
The force is always positive and it jumps when a core cluster attaches to a composite cluster. This corresponds
to changes of the color from orange to green. When the color changes from green to orange, a core cluster detaches from a composite cluster.
In such detachments events, the real soliton coordinate jumps by $\nb\sigma$. These jumps are not visible in the
continuous-coordinate representation in Fig.~\ref{fig:soliton-potential}(b). 

\subsection{Soliton velocity}
\label{subsec:soliton-velocity}
The time period of the soliton motion is
\begin{align}
\tau_{\rm sol}(y_1,y_2)&=\tau_{\nc}(y_1+\nb\sigma,y_2)\nonumber\\
&\phantom{=}{}+\tau_{\ncb}(y_2,y_1+\mb)\,,
\label{eq:tausol}
\end{align}
where $\tau_n(y,y')$ is the time for an $n$-cluster to move from $y$ to $y'$ given in Eq.~\eqref{eq:taun}.

The position $y_1$ is determined by the requirement that the $\nc$-cluster detaches from the $(\nc+\nb)$-cluster, 
i.e.\ the nonsplitting conditions~\eqref{eq:nonsplitting-conditions} for the composite $n=(\nc+\nb)$-cluster 
are violated for $l=\nb$:
\begin{equation}
\frac{1}{\nb}\sum\limits_{j=0}^{\nb-1}F(y_1+j\sigma)=
\frac{1}{\nc}\sum\limits_{j=\nb}^{\nb+\nc-1}F(y_1+j\sigma)\,.
\label{eq:y1-determ}
\end{equation}
If the system size $L$ is large enough (limit $L\to\infty$), the $\nb$-clusters have enough time to relax to their positions of mechanical equilibria. One can then 
set $y_2+\nc\sigma$ approximately equal to 
$x_{\nb}^\pm(\sigma,f)+j$ from Eq.~\eqref{eq:x0pm-finite-f}, with $n=\nb$ and $j$ the smallest integer satisfying 
$x_{\nb}^\pm(\sigma,f)+j>y_1+\nb\sigma$. This gives
\begin{equation}
y_2^\infty=x_{\nb}^\pm(\sigma,f)+j-\nc\sigma\,.
\label{eq:y2inf}
\end{equation}

The mean soliton velocity is the distance $\mb$ travelled in a time period $\tau_{\rm sol}$ 
divided by $\tau_{\rm sol}$,
\begin{equation}
v_{\rm sol}^\infty=\frac{\mb}{\tau_{\rm sol}(y_1,y_2^\infty)}\,.
\label{eq:vsolinfty}
\end{equation}
The corresponding expression for the type B mode of soliton motion is given in \ref{app:vsol-type2-solitons}.

An accurate treatment requires to 
consider the relaxation of the $\nb$-clusters towards positions of stable mechanical equilibria.
This will be discussed further below
in Sec.~\ref{sec:soliton-interaction} in connection with an effective soliton-soliton interaction. 

\section{Number of solitons}
\label{sec:number-of-solitons}
If $\Delta_{\rm min}(\sigma,L)$ fulfills the inequality \eqref{eq:Deltamin-selfconsistency}, solitons can 
occur for overfillings up to a maximal value satisfying $(\Delta_{\rm max}+L)\sigma< L$, i.e.\ 
\begin{equation}
\Delta_{\rm max}=\left\lfloor  \frac{(1-\sigma)L}{\sigma}\right\rfloor\,.
\label{eq:Deltamax}
\end{equation}
The number $N_{\rm sol}(\Delta)$  of solitons increases with the overfilling 
$\Delta$ and the minimal number $N_{\rm sol}^{\rm min}=N_{\rm sol}(\Delta_{\rm min})$
can be larger than one.

For determining $N_{\rm sol}^{\rm min}$, we consider the presoliton state 
with $\Delta=\Delta_{\rm min}\!-\!1$. 
It is composed of $N_{\rm b}(\Delta_{\rm min}\!-\!1)=N_{\rm pre}^{\rm b}$ clusters of the basic size $\nb$ 
and $n_{\rm res}$ residual particles. By adding one particle, 
the presoliton state rearranges into a nonequilibrium steady 
state, where a maximal number of stable $\nb$-clusters remains present.
Differently speaking, this state carries a minimal number $N_{\rm sol}^{\rm min}$ 
of running solitons 
and an integer number $N_{\rm b}(\Delta_{\rm min})<N_{\rm b}(\Delta_{\rm min}\!-\!1)$ of clusters of size $\nb$.
As described in Sec.~\ref{sec:soliton-propagation}, all
 clusters involved in the soliton propagation are formed out of the basic and the core cluster, 
i.e.\ we need the particles of one
basic stable cluster and the particles of one core cluster to generate one soliton. This means that 
\begin{equation}
\nbc=\nb+\nc
\label{eq:nbc}
\end{equation}
defines a soliton size in terms of number of particles.
We thus have
\begin{equation}
N_{\rm b}(\Delta_{\rm min}\!-\!1)\nb+n_{\rm res}+1=N_{\rm b}(\Delta_{\rm min})\nb+N_{\rm sol}^{\rm min}\nbc
\label{eq:Nsolmin-deltaNcl0-det1}
\end{equation}
or 
\begin{equation}
\nbc N_{\rm sol}^{\rm min}-\nb\delta N_{\rm b}^{(0)}=n_{\rm res}+1\,,
\label{eq:Nsolmin-deltaNcl0-det2}
\end{equation}
where $\delta N_{\rm b}^{(0)}=N_{\rm b}(\Delta_{\rm min}\!-\!1)-N_{\rm b}(\Delta_{\rm min})>0$
is the loss of $\nb$-clusters due to the formation of solitons at minimal overfilling $\Delta_{\rm min}$.

Equation~\eqref{eq:Nsolmin-deltaNcl0-det2} is a linear Diophantine equation 
in the two variables $N_{\rm sol}^{\rm min}$ and 
$\delta N_{\rm b}^{(0)}$. Since $\nb$ and $\nbc$ are coprime, 
it has the general solution \cite{Vorobyov:1980}
\begin{align}
N_{\rm sol}^{\rm min}&=-(n_{\rm res}\!+\!1)\frac{\nb^{\varphi(\nbc)}\!-\!1}{\nbc}+\nb j\,,
\label{eq:Nsolmin-solution-general}\\
\delta N_{\rm b}^{(0)}&=-(n_{\rm res}\!+\!1)\nb^{\varphi(\nbc)-1}+\nbc j\,,
\label{eq:deltaNcl0-solution-general}
\end{align}
where $j\in\mathbb{Z}$. Since $N_{\rm sol}^{\rm min}$ must be the smallest positive integer, 
$j=\lceil (n_{\rm res}\!+\!1)\nb^{\varphi(\nbc)-1}/\nbc\rceil$. Accordingly,
\begin{align}
N_{\rm sol}^{\rm min}&=\nb\left\lceil\frac{(n_{\rm res}\!+\!1)\nb^{\varphi(\nbc)-1}}{\nbc}\right\rceil
-(n_{\rm res}\!+\!1)\frac{\nb^{\varphi(\nbc)}\!-\!1}{\nbc}\,,
\label{eq:Nsolmin}\\
\delta N_{\rm b}^{(0)}&=\nbc\left\lceil\frac{(n_{\rm res}\!+\!1)\nb^{\varphi(\nbc)-1}}{\nbc}\right\rceil
-(n_{\rm res}\!+\!1)\nb^{\varphi(\nbc)-1}\,.
\label{eq:deltaNb0}
\end{align}

If $\Delta_{\rm max}>\Delta_{\rm min}$, we need to determine
$N_{\rm sol}(\Delta)$ also for 
$\Delta_{\rm min}<\Delta\le \Delta_{\rm max}$. To do this, we consider
the state for overfilling $(\Delta\!-\!1)$, which is composed of $N_{\rm sol}(\Delta\!-\!1)$ solitons of size $\nbc$ and 
$N_{\rm b}(\Delta\!-\!1)$ clusters of size $\nb$. Adding one particle, a state with 
$N_{\rm sol}(\Delta)$ solitons and $N_{\rm b}(\Delta)$ $\nb$-clusters is obtained, i.e.\ it must hold
\begin{equation}
N_{\rm sol}(\Delta\!-\!1)\nbc+N_{\rm b}(\Delta\!-\!1)\nb+1=N_{\rm sol}(\Delta)\nbc+N_{\rm b}(\Delta)\nb\,,
\label{eq:deltaNsol-det1}
\end{equation}
or
\begin{equation}
\nbc\delta N_{\rm sol}(\Delta)-\nb\delta N_{\rm b}(\Delta)=1\,,
\label{eq:deltaNcl-deltaNsol-det2}
\end{equation}
where $\delta N_{\rm sol}(\Delta)=N_{\rm sol}(\Delta)-N_{\rm sol}(\Delta\!-\!1)>0$ and 
$\delta N_{\rm b}(\Delta)=N_{\rm b}(\Delta\!-\!1)-N_{\rm b}(\Delta)>0$.

Equation~\eqref{eq:deltaNcl-deltaNsol-det2} has solutions only if $\nbc$ and $\nb$ are coprime.
It is a linear Diophantine equation 
in the two variables $\delta N_{\rm sol}$ and $\delta N_{\rm b}$, which
can be solved analogously to Eq.~\eqref{eq:Nsolmin-deltaNcl0-det2}. The solution is
\begin{align}
\delta N_{\rm sol}&=\nb\left\lceil\frac{\nb^{\varphi(\nbc)-1}}{\nbc}\right\rceil
-\frac{\nb^{\varphi(\nbc)}\!-\!1}{\nbc}\,,
\label{eq:deltaNsol-solution}\\[1ex]
\delta N_{\rm b}&=\nbc\left\lceil\frac{\nb^{\varphi(\nbc)-1}}{\nbc}\right\rceil
-\nb^{\varphi(\nbc)-1}\,.
\label{eq:deltaNcl-solution}
\end{align}
For $\Delta_{\rm min}$ not satisfying condition~\eqref{eq:Deltamin-selfconsistency} 
or $\Delta_{\rm max}=\Delta_{\rm min}$,
$\delta N_{\rm b}=\delta N_{\rm sol}=0$.

The solutions $\delta N_{\rm sol}$ and $\delta N_{\rm b}$ are independent of $\Delta$: with each additional
particle, the gain of solitons and the loss of stable clusters are the same.
It thus follows
\begin{align}
N_{\rm sol}(\Delta)&=N_{\rm sol}^{\rm min}+(\Delta-\Delta_{\rm min})\delta N_{\rm sol}\,,
\label{eq:Nsol}\\
N_{\rm b}(\Delta)&=N_{\rm pre}^{\rm b}\!-\!\delta N^{(0)}-(\Delta-\Delta_{\rm min})\delta N_{\rm b}\,.
\label{eq:Nb}
\end{align}
The $N_{\rm b}$ basic stable clusters and the $N_{\rm sol}$ soliton clusters fill the system, i.e.\ their
accommodating wells sum up to all $L$ potential wells:
\begin{equation}
N_{\rm b}\mb+N_{\rm sol}\mbc=L\,.
\label{eq:filling}
\end{equation}

We have checked the results for $N_{\rm sol}$ and $N_{\rm b}$ 
by simulations.

As an example, let us consider the particle diameter $\sigma_{p,q}=3/5$ and system length $L=23$, where $\nb=3$, 
$\Delta_{\rm min}=12$ and $\Delta_{\rm max}=15$ from Eq.~\eqref{eq:Deltamax}.
Equation~\eqref{eq:deltaNb0} gives $\delta N_{\rm b}^{(0)}=1$ and 
Eq.~\eqref{eq:Nsolmin} tells us that $N_{\rm sol}^{\rm min}=1$ soliton appears for the minimal overfilling 
$\Delta=\Delta_{\rm min}$. Equation~\eqref{eq:deltaNcl-solution} yields $\delta N_{\rm b}=3$, and 
Eq.~\eqref{eq:deltaNsol-solution} $\delta N_{\rm sol}=2$. This means that when 
increasing the overfilling in steps of one, the number of stable clusters decreases by three and 
the number of solitons increases by two.
At the maximal possible overfilling $\Delta_{\rm max}$, the number of solitons is 
$N_{\rm sol}(\Delta_{\rm max})=7$ according to Eq.~\eqref{eq:Nsol}.

%%%%%%%%%%%%%%%%%%%%%%%%%%%%%%%%%%%%%%%%%%%%%%%
\begin{figure}[t!]
\centering
\includegraphics[width=\columnwidth]{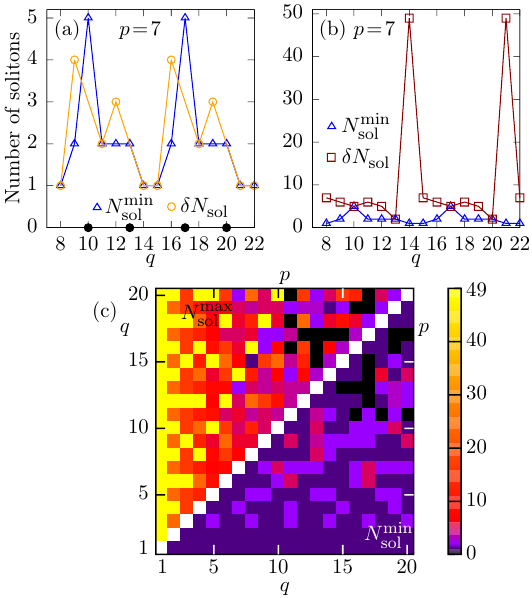}
\caption{(a) Number of solitons $N_{\rm sol}^{\rm min}$ 
at minimal overfilling $\Delta_{\rm min}$ 
[Eq.~\eqref{eq:Nsolmin}, blue triangles] and increment $\delta N_{\rm sol}$  
[Eq.~\eqref{eq:deltaNsol-solution}, orange circles] for particle diameters $\sigma_{p,q}=p/q$ 
and $p=7$ fixed.
(b) Minimal and maximal soliton numbers $N_{\rm sol}^{\rm min}$ (blue triangles) and $N_{\rm sol}^{\rm max}$ (red squares)
for the same particle diameters.
(c) Minimal and maximal soliton numbers $N_{\rm sol}^{\rm min}$ and $N_{\rm sol}^{\rm max}$ in dependence of $p$ and $q$
in a color-coded representation. Below (above) the diagonal $p=q$, $N_{\rm sol}^{\rm min}$ ($N_{\rm sol}^{\rm max}$) are plotted. In all panels, data are shown for a system with $L=50$ potential wells. Black symbols in (a) and (c) mark
particle diameters where no solitons form.}
\label{fig:minimal_and_maximal_soliton_number}
\label{fig:soliton_numbers}
\end{figure}
%%%%%%%%%%%%%%%%%%%%%%%%%%%%%%%%%%%%%%%%%%%%%%%

If the same particle diameter $\sigma_{p,q}=3/5$ is considered but a smaller system length $L=20$, $\nb=3$ and 
$\delta N_{\rm sol}=2$ remain unchanged as they are independent of $L$, 
while $\Delta_{\rm min}=11$ and $\Delta_{\rm max}=13$. 
Equations~\eqref{eq:Nsolmin}, \eqref{eq:deltaNb0} then yield 
$N_{\rm sol}^{\rm min}=2$, and Eq.~\eqref{eq:Nsol}
gives $N_{\rm sol}(\Delta_{\rm max})=6$. 

Figure~\ref{fig:soliton_numbers}(a)-(c) give soliton numbers in a system with
$L=50$ potential wells. In
Fig.~\ref{fig:soliton_numbers}(a), $N_{\rm sol}^{\rm min}$ (blue triangles) and
$\delta N_{\rm sol}$ (orange circles) are shown for different particle diameters $\sigma_{p,q}=p/q$ with $p=7$ fixed.
Both $N_{\rm sol}^{\rm min}$ and $\delta N_{\rm sol}$ are $p$-periodic functions of $q$.
Full black circles indicate particle sizes $\sigma_{p,q}$ where no solitons form,
because the self-consistency condition \eqref{eq:Deltamin-selfconsistency}
is violated. 

Figure~\ref{fig:soliton_numbers}(b) displays the minimal and maximal number $N_{\rm sol}^{\rm min}$ (blue triangles) and
$N_{\rm sol}^{\rm max}$ (red squares) for the same particle diameters as in Fig.~\ref{fig:soliton_numbers}(a).
$N_{\rm sol}^{\rm max}$ is a $p$-periodic function of $q$ also.
The large $N_{\rm sol}^{\rm max}$ at $q=mp$ can be explained as follows. 
The space $l_{q,p}$ covered by a soliton of size $q$ is 
$l_{q,p}=q\sigma_{p,q}=p/\gcd(p,q)$, 
where $\gcd(q,p)$ is the greatest common divisor of $q$ and $p$ 
(needed here because we vary $q$ at fixed $p$, implying that $q$ and $p$ are not always coprime). 
For $q=mp$, the minimal coverage $l_{q,p}=1$ is obtained, 
while for values $q$ different 
from integer multiples of $p$, $l_{q,p}$ is in general significantly larger. 
For example, $l_{q,p}=p$ if $q$ and $p$ are coprime.
For $\sigma_{q,p}=p/(mp)=1/m$, where $l_{q,p}=1$, $(L\!-\!1)$ solitons can appear
before the system would be fully covered by $L$ particles when adding one further particle.
Indeed, $N_{\rm sol}^{\rm max}=L\!-\!1=49$ at $q=mp$ in Fig.~\ref{fig:soliton_numbers}(b).

A color-coded representation of the minimal and maximal number 
of solitons in dependence of $p$ and $q$ is shown in 
Fig.~\ref{fig:soliton_numbers}(c), where below (above) the diagonal $p=q$ we give
$N_{\rm sol}^{\rm min}$ ($N_{\rm sol}^{\rm max}$). 

\section{Magic particle sizes}
\label{sec:magic-sizes}
Up to now, we have illustrated in the figures main features of
our analytical results.  Let us demonstrate, how these results can be applied to experiments, as, 
for example, the ones performed in \cite{Cereceda-Lopez/etal:2023},
with ratios of particle diameter to wavelength between 0.5 and 0.9. We may ask: if $0.5\le \sigma\le 0.9$ and say $L=20$,
around which magic particle diameters $\sigma_{p,q}=p/q$
solitons will appear for a given overfilling $\Delta$? And how many solitons are forming then?

These questions are answered as follows: For each $\sigma_{p,q}$, there is a maximal overfilling due to the condition
$N\sigma_{p,q}=(L+\Delta)\sigma_{p,q}<L$, i.e.\ $\Delta<(1-\sigma_{p,q})L/\sigma$. 
For example, $\Delta<20$ for $\sigma=1/2$, and $\Delta<20/9$ for $\sigma=9/10$.
This implies that solitons do not occur for all rational numbers $p/q$. To find those $\sigma_{p,q}$, for which solitons form,
one needs to check a limited range of $p$ and $q$ values only. This is because
$\nb\le(q-1)$ and $\nb\le N=L+\Delta$, i.e.\ the maximal possible $q$ value 
is $L+\Delta-1$. It implies that for given $\Delta$, only $q=2,\ldots,L+\Delta-1$ is possible and
for each of these $q$, the range of $p$ values is $1/2\le p/q\le 9/10$.
Using Eq.~\eqref{eq:Deltamin-determ},
$\Delta_{\rm min}(\sigma_{p,q},L=20)$ is calculated for each of the possible $p/q$.
If $\Delta\ge\Delta_{\rm min}$, solitons occur 
at $\sigma_{p,q}=p/q$.

\begin{figure}[b!]
\centering
\includegraphics[width=0.85\columnwidth]{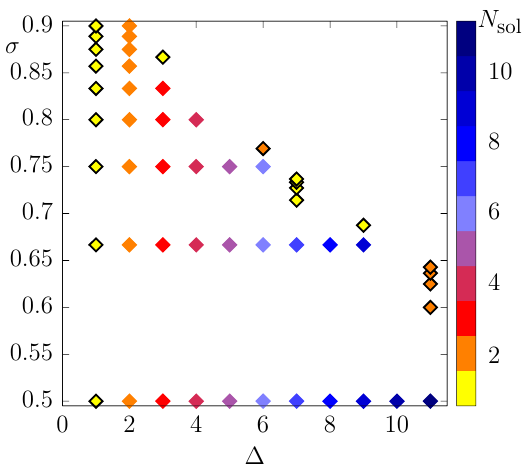}
\caption{Magic particle diameters in the range $0.5\le\sigma\le0.9$ around which solitons are predicted to 
form in a system with $L=20$ 
potential wells at weak drag forces. 
The number of solitons forming in the system is given by the color bar.}
\label{fig:magic_particle-sizes}
\end{figure}

Figure~\ref{fig:magic_particle-sizes} shows the results of this analysis. 
In addition, we have indicated by the color coding how
many solitons form according to Eq.~\eqref{eq:Nsol}. We suggest to test these predictions in experiments.

%%%%%%%%%%%%%%%%%%%%%%%%%%%%%%%%%%%%%%%%%%%%%%%
%%%%%%%%%%%%%%%%%%%%%%%%%%%%%%%%%%%%%%%%%%%%%%%

\section{Effective soliton-soliton interaction}
\label{sec:soliton-interaction}
In the presence of more than one soliton,
solitons can influence each other during their motion. 
Results of experiments and corresponding simulations reported in Ref.~\cite{Cereceda-Lopez/etal:2023} 
indeed indicate the existence of an effective repulsive soliton-soliton interaction. 

%%%%%%%%%%%%%%%%%%%%%%%%%%%%%%%%%%%%%%%%%%%%%%%
\begin{figure}[b!]
\centering
\includegraphics[width=0.8\columnwidth]{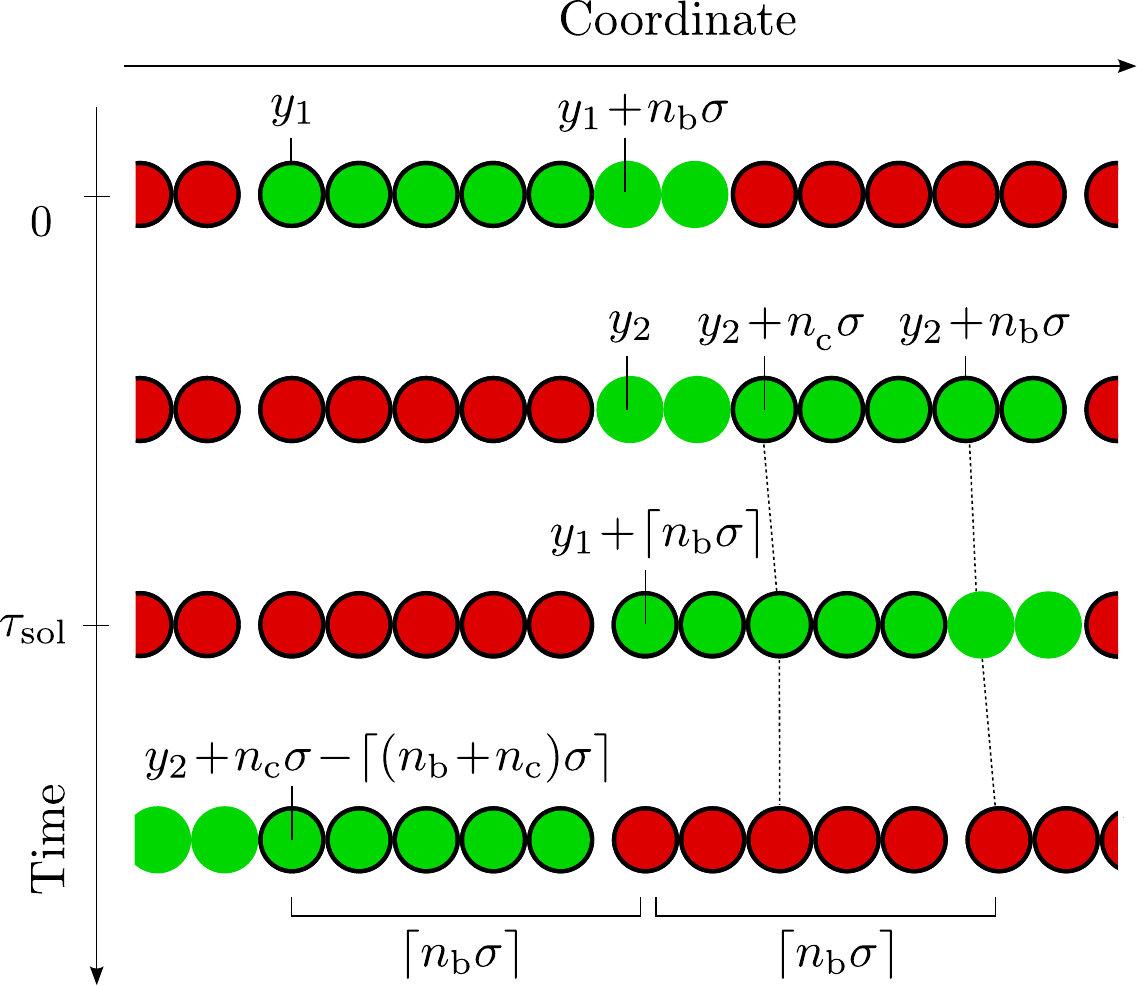}
\caption{Particle configurations replotted from Fig.~\ref{fig:illustration-soliton-propagation-modes}(a) to illustrate 
the effect of relaxation of the $\nb$-cluster on the soliton velocity.
In the particle configurations in the first line, the $\nc$-cluster detaches from a
composite $(\nb+\nc)$-cluster at position $y_1$ at time $t=0$. In the second line, the 
$\nc$-cluster attaches to the $\nb$-cluster at position $y_2$, and in the third line one period of the soliton motion is completed.
In the fourth line, not shown in Fig.~\ref{fig:illustration-soliton-propagation-modes}(a), the single soliton
has moved nearly through the whole system and reached a position, where the soliton
$\nc$-cluster attaches to the relaxing $\nb$-cluster initially at position $y_1$.}
\label{fig:illustration-for_y2_self-consitency-equation}
\end{figure}
%%%%%%%%%%%%%%%%%%%%%%%%%%%%%%%%%%%%%%%%%%%%%%%

We here show that this effective repulsive interaction is related to the relaxation of the $\nb$-clusters
towards their positions of mechanical equilibria. The effect occurs already in a small system 
with a single soliton and manifests
itself in a slowing down of the soliton velocity upon decreasing the system size $L$.

To illustrate our derivation of this slowing down, we replot in the first 
three lines of Fig.~\ref{fig:illustration-for_y2_self-consitency-equation} 
the configurations shown in the first, third and last line of 
Fig.~\ref{fig:illustration-soliton-propagation-modes}(a).
In the additional configuration shown in the fourth line of Fig.~\ref{fig:illustration-for_y2_self-consitency-equation}, 
the single soliton has propagated nearly through the whole system and reached a position, where the
soliton $\nc$-cluster attaches to the relaxing $\nb$-cluster initially at position $y_1$. At that moment, the soliton $\nc$-cluster 
has a position $y_2'$, which can be determined by comparing the configurations in the second and fourth line.
The soliton has moved  $N_{\rm b}$ periods between the two configurations, i.e.\ traveled a distance $N_{\rm b}\mb$. 
Because $N_{\rm b}\mb+\mbc=L$ for a single soliton, see Eq.~\eqref{eq:filling}, we can write $N_{\rm b}\mb=L-\mbc$ 
for this distance, and $y_2'=(y_2+L-\mbc)\ourmod L=y_2-\mbc$ (we can assume $y_2>\mb$). 
The $\nb$-cluster, to which the soliton $\nc$-cluster attaches, has position $y_2'+\nc\sigma=y_2-\mbc+\nc\sigma$.

In the steady state, the time for the $\nb$-cluster to relax from position $y_1$ to
$y_2-\mbc+\nc\sigma$ 
must be equal to the time of the initially detaching $\nc$-cluster to move from $y_1+\nb\sigma$ to $y_2$ plus the time 
$(L-\mbc)/v_{\rm sol}=(L-\mbc\tau_{\rm sol}/\mb$ 
for the soliton to move the distance $L-\mbc$:
\begin{align}
&\tau_{\nb}(y_1,y_2-\mbc+\nc\sigma)\label{eq:y2-determ}\\
&\hspace{2em}=\tau_{\nc}(y_1+\nb\sigma,y_2)+\frac{L-\mbc}{\mb}\,\tau_{\rm sol}(y_1,y_2)\,.
\nonumber
\end{align}
Here, $\tau_{\rm sol}(y_1,y_2)$ and $y_1$ are given by Eqs.~\eqref{eq:tausol} and \eqref{eq:y1-determ}. 
Equation~\eqref{eq:y2-determ}
is a self-consistency equation for $y_2$, whose solution depends on $L$, $y_2=y_2(L)$.

Inserting $y_1$ and $y_2(L)$ into Eq.~\eqref{eq:tausol} gives the
relaxation-corrected soliton period $\tau_{\rm sol}(y_1,y_2(L))$ and the corresponding
soliton velocity 
\begin{equation}
v_{\rm sol}(1,L)=\frac{\mb}{\tau_{\rm sol}(y_1,y_2(L))}
\label{eq:vsol}
\end{equation}
for a single soliton ($N_{\rm sol}=1$). 
The limit $L\to\infty$ corresponds to the approximation when disregarding the relaxation of the $\nb$-clusters, i.e.\
$v_{\rm sol}(1,L\to\infty)=v_{\rm sol}^\infty=\mb/\tau_{\rm sol}(y_1,y_2^\infty)$. For type B solitons,
the expression for the relaxation-corrected velocity can be derived analogously, see \ref{app:vsol-type2-solitons}.

\begin{figure}[t!]
\centering
\includegraphics[width=\columnwidth]{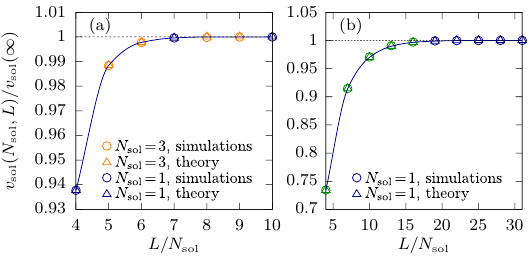}
\caption{Velocity of a soliton as a function of the mean distance $L/N_{\rm sol}$ between solitons 
when taking into account the relaxation of the basic stable $\nb$-clusters towards positions of mechanical equilibria. 
The velocity is normalized with respect to its value for $L\to\infty$ and
results are shown for particle diameter
$\sigma=0.57$, and drag forces (a) $f=0.05$ and (b) $f=0.15$. 
For a given system size $L$, the velocity was calculated for the running state with minimal overfilling 
$\Delta=\Delta_{\rm min}$.
Circles mark simulation results and triangles the theoretical 
results for the relaxation-corrected soliton velocity based on Eq.~\eqref{eq:vsol-multiple-solitons}.
Blue symbols mark states with one soliton of type A, orange symbols with three solitons of type A, and
green symbols in (b) with one soliton of type B (occurring for $L/N_{\rm sol}\le 16$).
The lines are guides for the eye, connecting the triangles.}
\label{fig:vsol(L)}
\end{figure}

The relaxation-corrected $v_{\rm sol}(1,L)$ decreases when $L$ 
becomes smaller, and the slowing down of the soliton motion
is more pronounced and extends to larger $L$ when $f$ becomes larger. 
This is demonstrated by the single-soliton $N_{\rm sol}=1$ data in Fig.~\ref{fig:vsol(L)}, where we show 
$v_{\rm sol}(1,L)/v_{\rm sol}^\infty$ for $\sigma=0.57$, and two drag forces
$f=0.05$ and $f=0.15$ at minimal overfilling $\Delta=\Delta_{\rm min}$.

We can say that in the single-soliton state, the soliton interacts effectively with itself at a distance $L$. 
In the presence of several solitons, the distance between neighboring solitons plays the role of $L$. The 
slowing down of the motion can be interpreted as a repulsive force that lets the solitons to stay apart from each other.

%%%%%%%%%%%%%%%%%%%%%%%%%%%%%%%%%%%%%%%%%%%%%%%
%%%%%%%%%%%%%%%%%%%%%%%%%%%%%%%%%%%%%%%%%%%%%%%	

For $N_{\rm sol}>1$, the velocity $v_{\rm sol}(N_{\rm sol},L)$
of each soliton should be the relaxation-corrected single-soliton velocity $v_{\rm sol}(1,L)$
evaluated at the mean distance $L/N_{\rm sol}$ between the solitons, i.e.\ 
\begin{equation}
v_{\rm sol}(N_{\rm sol},L)=v_{\rm sol}(1,L/N_{\rm sol})\,.
\label{eq:vsol-multiple-solitons}
\end{equation}
The multiple soliton data ($N_{\rm sol}=3$) in Fig.~\ref{fig:vsol(L)} 
are in excellent agreement with this prediction.

\section{Soliton-mediated particle currents}
\label{sec:particle_currents}
For calculating the mean particle current mediated by a single soliton, we use the following concept: when starting with
a certain particle configuration, an equivalent configuration occurs after a 
minimal number $k$ of circulations of the soliton around the system. In the equivalent configuration,
all clusters are formed by the same particles as in the initial configuration and the soliton's position with respect to the particles in the $\nb$-clusters is the same. The only difference is that in the equivalent configuration the particles are displaced
relative to those in the initial configuration. Because of the periodicity of the dynamics,
the respective displacement $d$ must be the same for all particles and equal to an integer number of potential wells.

The soliton thus has travelled a distance $kL+d$ when the equivalent configuration occurs, and the time needed for moving
this distance is  $\Delta t=(kL+d)/v_{\rm sol}(1,L)$. The mean velocity of each particle is $v_{\rm p}=d/\Delta t$
in the stationary state and the mean particle current is $j_{\rm st}=(N/L)v_{\rm p}$, i.e.\
\begin{equation}
j_{\rm st}(1,L)=\frac{d}{kL+d}\frac{N}{L}v_{\rm sol}(1,L)\,.
\label{eq:jst-single-soliton}
\end{equation}

In the presence of $N_{\rm sol}>1$ solitons, $k$ circulations of one soliton imply that all solitons have circulated $k$ times. Because $k$ circulations of one soliton displace particles by $d$, $N_{\rm sol}$ solitons displace them by $N_{\rm sol} d$. 
The particle current thus follows from Eq.~\eqref{eq:jst-single-soliton} when replacing $d$ with $N_{\rm sol}d$,
\begin{equation}
j _{\rm st}(N_{\rm sol},L)=\frac{N_{\rm sol} d}{k L+ N_{\rm sol} d}\frac{N}{L}\,v_{\rm sol}(N_{\rm sol},L)\,.
\label{eq:jst}
\end{equation}
It remains to determine $k$ and $d$ for a system carrying a single soliton.

To derive $k$, we analyze how the order number $s$ of a tagged particle in an $\nb$-cluster changes after each soliton circulation. We define the order number of a particle in a cluster as its location within the cluster minus one, i.e.\ 
$s=0$ for the first particle in the cluster, $s=1$ for the second particle and so on.
When the $\nc$-cluster attaches to the considered $\nb$-cluster,
the original order number $s$ of the tagged particle in the $\nb$-cluster changes to $s+\nc$
in the composite $(\ncb)$-cluster.
After the soliton passage, or after one circulation of the soliton, the tagged particle is part of an $\nb$-cluster 
and its shifted order number $s+\nc$ has to be taken modulo $\nb$, i.e.\ the order number $s^{(1)}$ after one soliton circulation is 
\begin{equation}
s^{(1)}=(s+\nc)\ourmod\nb\,.
\label{eq:s1}
\end{equation}

Depending of the value of $(s+\nc)$, this is equal to either
$s_1^{(1)}=s+\nc\ourmod\nb$ or to $s_2^{(1)}=s+\nc\ourmod\nb-\nb$. Accordingly, there are only two possible values 
$\Delta s_{1,2}$ for the change of order number $\Delta s$ after one soliton circulation:
\begin{align}
\Delta s_1&=\nc\ourmod\nb>0\,,\label{eq:s1}\\
\Delta s_2&=\nc\ourmod\nb-\nb<0\,.\label{eq:s2}
\end{align}
That only two values are possible is due to the fact that for $\Delta s_1>1$
the particles of an $\nb$-cluster are divided into two sets after a soliton passage, where in each set
they have the same ordering as before and where one set forms the front part and the other the back part
of two different $\nb$-clusters after the passage.

If $\Delta s_1=0$, the same ordering of the particles is obtained
already after one soliton circulation. In that case, $k=1$ and $\Delta s_2$ has no meaning.

After two soliton circulations, the order number is 
$s^{(2)}\!=\!(s^{(1)}+\nc)\ourmod\nb\!=\![((s+\nc)\ourmod\nb)\!+\!\nc]\ourmod\nb\!=\!(s\!+\!2\nc)\ourmod\nb$,
and after $j$ soliton circulations it is $s^{(j)}=(s+j\nc)\ourmod\nb$. 
After $k$ circulations, the order number must be the same as initially, i.e.\ we obtain
$s^{(k)}=(s+k\nc)\ourmod\nb=s$ as determining equation for $k$. 
This gives $k\nc=l\nb$ with $l\in\mathbb{N}$. After division by $\gcd(\nb,\nc)$ this becomes
$k\nc'=l\nb'$ where $\nc'=\nc/\gcd(\nb,\nc)$ and $\nb'=\nb/\gcd(\nb,\nc)$.
The smallest positive $k$ and $l$ solving this equation are $k=\nb'$ and $l=\nc'$, yielding
\begin{equation}
k=\frac{\nb}{\gcd(\nb,\nc)}\,.
\label{eq:k-circulations}
\end{equation}
Note that if $\Delta s_1=\nc\ourmod\nb=0$, $\nc$ is divisible by $\nb$ and accordingly $k=1$, in agreement with
the discussion after Eqs.~\eqref{eq:s1}, \eqref{eq:s2} above. We note that $k=1$ is possible only for $\nb=1$, because
$\nb$ and $(\nb+\nc)$ are coprime [see Eq.~\eqref{eq:deltaNcl-deltaNsol-det2}].

The derivation of $d$ is a bit more involved and given in \ref{app:d-displacement}. The result is
\begin{align}
d&=\frac{\nb\!-\!\nc\ourmod\nb}{\gcd(\nb,\nc)}l_1\mb+\frac{\nc\ourmod\nb}{\gcd(\nb,\nc)}l_2\mb\nonumber\\[0.5ex]
&\hspace{1em}{}-k\mbc
\label{eq:d}
\end{align}
where
\begin{equation}
l_\alpha=1+\left\lfloor \frac{\mbc-\Delta s_\alpha\sigma}{\mb}\right\rfloor\,,\hspace{1em}\alpha=1,2\,.
\label{eq:lalpha}
\end{equation}

\begin{figure}[t!]
\centering
\includegraphics[width=0.8\columnwidth]{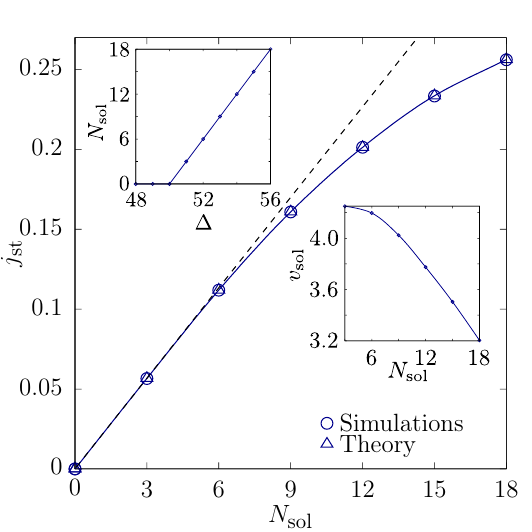}
\caption{Simulated mean particle current $j_{\rm st}$ in the stationary state (circles) as a function of overfilling $N_{\rm sol}$
in comparison with that calculated from Eq.~\eqref{eq:jst} (triangles). 
The system size is $L=75$. All other parameters are the same as in  Fig.~\ref{fig:vsol(L)}(b). 
The dashed line shows the linear increase of $j_{\rm st}$ with $N_{\rm sol}$,
if incomplete relaxations of $\nb$-clusters are neglected. The solid line is a guide for the eye. 
The upper inset gives the dependence of $N_{\rm sol}$ on the overfilling $\Delta$, and 
the lower inset the dependence of the single-soliton velocity on $N_{\rm sol}$.}
\label{fig:current_vs_overfilling}
\end{figure}

We have validated Eq.~\eqref{eq:jst} for various parameter sets.
As an example, we show in Fig.~\ref{fig:current_vs_overfilling} simulated particle currents and
$j_{\rm st}$ calculated from Eq.~\eqref{eq:jst} as a function of $N_{\rm sol}$ for
a system size $L=75$ and otherwise the same parameters as in Fig.~\ref{fig:vsol(L)}.
The number $N_{\rm sol}$ increases linearly with the overfilling $\Delta$ for $\Delta>\Delta_{\rm min}$, 
see the upper inset.
The decrease of the single-soliton velocity $v_{\rm sol}$ with $N_{\rm sol}$ shown in the other inset
is due to incomplete relaxation of $\nb$-clusters. It
leads to the sublinear increase of $j_{\rm st}$ with $N_{\rm sol}$ for $N_{\rm sol}>6$. The dashed line 
shows the behavior if the relaxation
of $\nb$-clusters is neglected.
The slowing-down of $v_{\rm sol}$ with $N_{\rm sol}$
corresponds to the effective repulsive soliton-soliton interaction discussed in Sec.~\ref{sec:soliton-interaction}. 

%%%%%%%%%%%%%%%%%%%%%%%%%%%%%%%%%%%%%%%%%%%%%%%
%%%%%%%%%%%%%%%%%%%%%%%%%%%%%%%%%%%%%%%%%%%%%%%

Equation~\eqref{eq:jst} is exact for a finite system of size $L$ under periodic boundary conditions. 
The result can be used to derive
the dependence of the stationary current on the particle density $\rho=N/L>1$
in the thermodynamic limit $L\to\infty$. To this end, we first note that $k$ and $d$ 
in Eqs.~\eqref{eq:k-circulations} and \eqref{eq:d}
are independent of $L$, because they are fully determined by $\nb$ and $\nc$.
The number $N_{\rm sol}$ at fixed $\rho$ should increase linearly with $L$ in the thermodynamic limit.
This is indeed the case: from Eq.~\eqref{eq:deltaNsol-solution}
we obtain $N_{\rm sol}\sim (\Delta-\Delta_{\rm min})\delta N_{\rm sol}\sim
(\rho-1-\Delta_{\rm min}/L)\delta N_{\rm sol}L$, where $\delta N_{\rm sol}$ 
is independent of $L$ [see Eq.~\eqref{eq:deltaNsol-solution}].
From Eq.~\eqref{eq:Deltamin-determ} follows 
\begin{equation}
\frac{\Delta_{\rm min}}{L}\sim \left(\frac{\nb}{\mb}-1\right)=(\rho_{\rm b}-1)\,,
\end{equation}
for $L\to\infty$, where 
\begin{equation}
\rho_{\rm b}=\frac{\nb}{\mb}\,.
\end{equation}
Accordingly, $N_{\rm sol}/L\sim (\rho-\rho_{\rm b})\delta N_{\rm sol}$ and 
because $N_{\rm sol}>0$, we have to require $\rho>\rho_{\rm b}$. This means that
$\rho_{\rm b}$ is a critical density: for $\rho<\rho_{\rm b}$, particle transport is governed by
thermally activated dynamics \cite{Lips/etal:2018}, while for $\rho>\rho_{\rm b}$, it is governed
by persistent solitons.
With  $v_{\rm sol}(N_{\rm sol},L)=v_{\rm sol}(1,L/N_{\rm sol})$,
Eq.~\eqref{eq:jst} yields the bulk current-density relation for soliton-mediated particle
transport ($\rho>\rho_{\rm b}$):
\begin{equation}
j_{\rm st}^{\rm b}(\rho)=\frac{\displaystyle\rho(\rho-\rho_{\rm b})}
{\displaystyle\frac{k}{d\,\delta N_{\rm sol}}+(\rho-\rho_{\rm b})}\,
v_{\rm sol}\left(1,\frac{1}{(\rho\!-\!\rho_{\rm b})\delta N_{\rm sol}}\right)\,.
\label{eq:jst-thermodynmic-limit}
\end{equation}

The main results of this section are Eq.~\eqref{eq:jst} for the soliton-mediated current in finite systems and
Eq.~\eqref{eq:jst-thermodynmic-limit} for the current-density relation
in the thermodynamic limit. The latter predicts a dynamical phase transition to occur 
from a jammed to a current-carrying state at a particle number density $\rho_{\rm b}=\nb/\mb$ 
for $k_{\rm B}T/U_0\ll1$, strictly speaking in the limit $k_{\rm B}T/U_0\to0$.
We believe that the strong increase of currents found in earlier investigations of current-density relations
at $k_{\rm B}T/U_0=1/6$ for $\rho>1$ \cite{Lips/etal:2019} is reflecting this transition.
It would be interesting to
see whether it can be obtained by continuum theories, as, e.g., the dynamical density functional 
theory \cite{teVrugt/etal:2020},
the power density functional theory \cite{Schmidt:2022, delasHeras/etal:2023}, 
or the macroscopic fluctuation theory \cite{Bertini/etal:2015}.

\section{Particle diameters larger than wavelength of potential}
\label{sec:sigma>1}
 So far we have considered a system of size $L$ with $N$ particles of diameters $0<\sigma<1$. 
The particle dynamics in a system of size $L'$ with $N'$ particles having diameter $\sigma'>1$ 
can be mapped onto that in a system
of size $L=L'-N\lfloor\sigma'\rfloor$ with the same number $N=N'$ of particles having diameter 
$\sigma=\sigma'-\lfloor\sigma'\rfloor<1$. 
More precisely, there exists a one-to-one correspondence of Brownian paths in the respective systems \cite{Lips/etal:2019}, 
implying that physical quantities and properties of solitons in both systems can be related to each other. 

As a consequence, neither the soliton cluster sizes $\nb$ and $\nc$ nor the numbers  $N_{\rm b}$ and $N_{\rm sol}$
of stable clusters and solitons change. 
Due to our definition of the overfilling $\Delta=N-L$ in the system with $\sigma<1$, the value in the corresponding system with $\sigma'>1$ 
is $\Delta'=\Delta-N\lfloor \sigma'\rfloor$. For $\sigma'>1$, however, the particles are not fitting into wells, implying that
$\Delta'$ does not have the meaning of an overfilling of the potential wells by particles. 

A bit more care has to be taken when transforming soliton velocities and particle currents. For the sake of brevity, we refer to the system with the primed quantities as the ``primed system''.
The soliton velocity in Eq.~\eqref{eq:vsolinfty} without relaxation correction becomes 
$v'_{\rm sol}=v_{\rm sol}+\nb\lfloor\sigma'\rfloor/\tau'_{\rm sol}$, where the time period of the soliton motion does not change, $\tau'_{\rm sol}=\tau_{\rm sol}$. The relaxation correction for the soliton velocity in the primed system can be 
determined analogously to that in the unprimed system as described in Sec.~\ref{sec:soliton-interaction}.  In Eq.~\eqref{eq:y2-determ}, 
one needs to replace $L$ and $\sigma$ by $L'$ and $\sigma'$,
and can take $y_1'=y_1$. It then becomes a determining for $y_2'(L')$. Replacing all unprimed by the primed quantities in Eqs.~\eqref{eq:tausol} and \eqref{eq:vsol} gives the relaxation-corrected single soliton velocity 
$v'_{\rm sol}(1,L')=\lceil \nb\sigma'\rceil/\tau_{\rm sol}(y_1,y_2'(L'))$. 
Using this in Eq.~\eqref{eq:vsol-multiple-solitons}, one obtains
$v_{\rm sol}'(N_{\rm sol},L')=v_{\rm sol}'(1,L'/N_{\rm sol})$. As for the particle current in Eq.~\eqref{eq:jst}, 
$d$ and $k$ as well as $N$ and $N_{\rm sol}$
are unchanged, i.e.\ one needs to replace $L$ by $L'$ and $v_{\rm sol}(N_{\rm sol},L)$ by $v_{\rm sol}'(N_{\rm sol},L')$.

\section{Conclusions}
\label{sec:conclusions}
We have studied the conditions for soliton appearance, soliton properties, and soliton-mediated particle currents
in Brownian transport of hard spheres 
through a sinusoidal potential
with potential barriers much larger than the thermal energy,  i.e.\ when effects of thermal noise are negligible.
The solitons manifest themselves as propagating waves of particle clusters, which are formed by
periodically repeating mergers and splittings of clusters. 

A sufficiently high number of particles is needed for the cluster waves to appear. 
The minimal number can be calculated from the presoliton state, which is the
mechanically stable state with largest particle number.
If a particle is added to the presoliton state, solitons occur and they start to propagate under the influence of an external driving. In this study we have considered a constant drag force $f$, which can be
experimentally realized in the comoving frame of colloidal particles driven 
by rotating optical traps \cite{Lutz/etal:2006, Tsuji/etal:2020, Tsuji/etal:2022, Cereceda-Lopez/etal:2023}, 
or by translation of optical traps with a constant velocity \cite{ Bohlein/etal:2012, Bohlein/Bechinger:2012}.

The presoliton state is formed by evenly 
spaced mechanically stable $\nb$-clusters,
i.e.\ clusters formed by $\nb$ particles in contact. 
Surprisingly, mechanical stabilizability of a cluster
is related to a purely geometric property,  namely 
the residual free space left in potential wells accommodating the cluster. As a consequence,
the number $\nb$ is determined by a principle of minimum free residual space. Using this principle,
we derived an explicit expression for $\nb$, see Eqs.~\eqref{eq:nb-solution}, \eqref{eq:nast-determ2}.

We believe that there exists a connection between mechanical stabilizability and
free residual space for all symmetric potentials with a single period, maybe with a generalization that
the presoliton state is formed by more than one basic cluster.
Generally, one may ask whether laws exist for largest mechanically stable packings of hard spheres in periodic potentials. 
This question can be viewed as an extension of the well-known closest packing problem of hard spheres in free space.

In the running state, the cluster sequence in the soliton propagation has a smallest core cluster made of $\nc$ particles. All 
other clusters involved are composed of one $\nc$-cluster and $\nb$-clusters. The size $\nc$ follows from the
requirement of barrier-free motions of  soliton clusters, which for infinitesimal weak forces 
is given by Eq.~\eqref{eq:nc-zero-force-limit}.
The simplest 
soliton propagation mode is that of alternating movements of an $\nc$- and a composite $(\ncb)$-cluster. 
It resembles the situation in a relay race: the $\nc$-cluster can by viewed as the relay that is carried by a core runner and
passed to another runner, when an $\nc$-cluster attaches to the left end of an $\nb$-cluster and an $(\nc+\nb)$-cluster forms. 
Subsequently the relay is passed to another 
core runner, when the $\nc$-cluster detaches from the right end of the $(\nc+\nb)$-cluster, and so on.

That the soliton propagation involves solely $\nb$- and $\nc$-clusters can be understood 
from a general argument: a presoliton state if formed essentially by a periodic spatial arrangement of 
basic $\nb$-clusters and one additional cluster is 
needed to bridge the gaps between the $\nb$-clusters. We thus expect such cluster waves to occur 
in general in periodic potentials.

The incomplete relaxation of $\nb$-clusters towards their positions of mechanical equilibria in a running state gives rise to 
an effective repulsive soliton-soliton interaction, which corresponds to a decrease of soliton velocities as described by
Eqs.~\eqref{eq:vsol} [or \eqref{eq:vsolB}] and \eqref{eq:vsol-multiple-solitons}. This effect, recently observed 
experimentally \cite{Cereceda-Lopez/etal:2023}, is also responsible for a slowing down of particle currents, 
see Fig.~\ref{fig:current_vs_overfilling} and Eqs.~\eqref{eq:jst}, \eqref{eq:jst-thermodynmic-limit}. 

In experiments like that performed in Ref.~\cite{Cereceda-Lopez/etal:2023},
key results of our theoretical analysis can be tested
by studying how the system's state changes
when incrementing the particle number at low temperatures.
For example, one can choose parameters as in Fig.~\ref{fig:illustration_open_problems}, i.e.
a ratio $\sigma/\lambda=0.6$ of particle diameter to wavelength of the potential, a system with
$L/\lambda=75$ potential wells,
and a weak drag force $f=0.1\fc\cong0.314U_0/\lambda$.
According to Eqs.~\eqref{eq:nb-solution} and \eqref{eq:Nbpre}, 
the presoliton state is formed by $N^{\rm b}_{\rm pre}=37$ 
basic mechanically stable clusters of size $\nb=3$ 
and one residual particle. Equation~\eqref{eq:Nsolmin} tells us
that $N_{\rm sol}^{\rm min}=1$ soliton forms at the minimal required overfilling $\Delta_{\rm min}$,
and Eq.~\eqref{eq:deltaNsol-solution} predicts that the number of solitons 
increases by $\Delta N_{\rm sol}=2$ with each further added particle until
the maximal overfilling $\Delta_{\rm max}=49$ [Eq.~\eqref{eq:Deltamax}] is reached.

In our previous work \cite{Antonov/etal:2022a}, restricted to overfilling $\Delta=1$, we showed that
soliton behavior at low temperatures allows one to understand particle transport 
at higher temperatures. An important effect 
is the thermal activation of transient solitons, which already occur for particle numbers  below
minimal overfilling.  We have reported first results on these thermally activated solitons for a system with filling 
factor one \cite{Antonov/etal:2022b} ($N=L$).
By identifying a transition state, we derived a soliton generation rate and by considering defects left after soliton 
generation, we determined soliton life times between generation and annihilation. Based on the generation rates and 
lifetimes, a scaling theory could be developed to describe how particle currents vary with 
particle diameter $\sigma$ and  system length $L$. It should be possible to extend this methodology 
to thermally activated solitons under conditions of overfilling ($N>L$). 

Knowledge of the properties of thermally activated solitons will be important in particular for studying the 
thermodynamic limit $L\to\infty$, where either the density $N/L$ of particles is kept fixed, 
or the overfilling $\Delta=N-L$. In the first case, the particle  dynamics should be 
governed by persistent solitons for large $L$, while in the second case 
by thermally activated solitons. 

We expect the theoretical concepts presented here
to be relevant also for particle transport under time-dependent driving and in dimensions higher than one. 
For the numerous examples given in the Introduction \cite{Xiao/etal:2003, Paneth:1950, Zepeda-Ruiz/etal:2004, Landau/etal:1993,Matsukawa/Zinkle:2007, vanderMeer/etal:2018, Korda/etal:2002, Bohlein/etal:2012, Bohlein/Bechinger:2012, Vanossi/etal:2012, Brazda/etal:2018, Vanossi/etal:2020, deSouzaSilva/etal:2006, Tierno/Fischer:2014, Stoop/etal:2020, Lips/etal:2021, Juniper/etal:2015}, as well as in new setups of twisted optical and magnetic lattices \cite{Stuhlmueller/etal:2022, Stuhlmueller/etal:2024}, they can trigger new ways for analyzing cluster dynamics on a microscopic level.

\section*{Acknowledgements}
Financial support by the Czech Science Foundation (Project No.\ 23-09074L) and the 
Deutsche Forschungsgemeinschaft (Project No.\ 521001072) is gratefully acknowledged.
We thank P. Tierno for very helpful discussions on the impact of our results for experiments. 

\appendix
\counterwithin{figure}{section}
\counterwithin{table}{section}

\section{Proof of free-space theorem}
\label{app:proof-free-space-theorem}
When inserting the cosine potential from Eq.~\eqref{eq:cosine-potential} into the non-splitting conditions~\eqref{eq:nonsplitting-conditions}
for a stabilizable $n$-cluster at position $x$, the inequalities become
\begin{align}
&\frac{1}{l}\sum\limits_{j=0}^{l-1}\!\sin[2\pi(x\!+\!j\sigma_{p,q})]-
\frac{1}{n\!-\!l}\sum\limits_{j=l}^{n-1}\!\sin[2\pi(x\!+\!j\sigma_{p,q})]\nonumber\\
&=\Biggl[\frac{\sin(\pi l\sigma_{p,q})}{l\sin(\pi\sigma_{p,q})}\sin[2\pi x+\pi(l\!-\!1)\sigma_{p,q}]
\label{eq:split2}\\
&\hspace{2em}{}-\frac{\sin[\pi (n\!-\!l)\sigma_{p,q}]}{(n\!-\!l)\sin(\pi\sigma_{p,q})}
\sin[2\pi x\!+\!\pi (n\!+\!l\!-\!1)\sigma_{p,q}]\Biggr]>0\,.
\nonumber
\end{align}
Inserting $x=x_n^+(\sigma_{p,q},0)$ from Eq.~\eqref{eq:x0pm-finite-f}, or any equivalent position shifted by an integer, we get
\begin{align}
&\Biggl[\frac{1}{l}\frac{\sin(\pi l\sigma_{p,q})}{\sin(\pi\sigma_{p,q})}\sin[\pi+\pi(l\!-\!n)\sigma_{p,q}]\\
&\hspace{2em}{}-\frac{1}{n-l}\frac{\sin[\pi (n\!-\!l)\sigma_{p,q}]}{\sin(\pi\sigma_{p,q})}\sin(\pi+\pi l\sigma_{p,q})\Biggr]>0\,.
\nonumber
\end{align}
Since $\sin(\pi\sigma_{p,q})>0$ for $\sigma_{p,q}<1$, this gives
\begin{equation}
\sin(\pi l\sigma_{p,q})\sin[\pi(\nb-l)\sigma_{p,q}]>0\,.
\end{equation}
Because $\sin[\pi(n\!-\!l)\sigma_{p,q}]>0$ for considered $x_n^+(\sigma_{p,q},0)$, 
see Eq.~\eqref{eq:x0pm-finite-f}, it further follows 
\begin{equation*}
[\sin(\pi n\sigma_{p,q})\cos(\pi l\sigma_{p,q})-\sin(\pi l\sigma_{p,q})\cos(\pi n\sigma_{p,q})]>0\,,
\end{equation*}
which, after
division by $\sin^2(\pi l\sigma_{p,q})\sin(\pi n\sigma_{p,q})>0$, yields
\begin{equation}
\cot(\pi l\sigma_{p,q})>\cot(\pi n\sigma_{p,q})\,.
\end{equation}
Because the $\cot$-function is $\pi$-periodic, this is equivalent to
\begin{equation}
\cot(\pi l\sigma_{p,q}\!-\!\pi\lceil l\sigma_{p,q}\rceil)>\cot(\pi n\sigma_{p,q}\!-\!\pi\lceil n\sigma_{p,q}\rceil)\,.
\label{eq:cot-inequility}
\end{equation}
The arguments of the $\cot$-functions in this relation lie in the interval 
$]-\pi,0[$ and the $\cot$-function is monotonically decreasing in each of its branches. We thus obtain from
\eqref{eq:cot-inequility}
\begin{equation}
\lceil n\sigma_{p,q}\rceil-n\sigma_{p,q}<
\lceil l\sigma_{p,q}\rceil-l\sigma_{p,q}\,,\quad l=1,\ldots,n\!-\!1\,.
\label{eq:consistency-inequality}
\end{equation}
These are the inequalities~\eqref{eq:free-space-theorem} of the free-space theorem.
The derivation for a position $x_n^-(\sigma,0)$ of the $n$-cluster is analogous. 

\section{Velocity of type B solitons}
\label{app:vsol-type2-solitons}
The time period for the mode B of soliton propagation is
\begin{align}
\tau_{\rm sol}^{\rm B}(y_1,y_2,y_3,y_4)&=\tau_{\nc}(y_1+\nb\sigma,y_2)\nonumber\\
&\hspace{-2em}{}+\tau_{\nc+\nb}(y_2,y_3+\nb\sigma)+\tau_{\nb+\nc+\nb}(y_3,y_4)\nonumber\\
&\hspace{1em}{}+\tau_{\nc+\nb}(y_4,y_1+\mb)\,,
\label{eq:tausolB}
\end{align}
where $y_1,\ldots,y_4$ refer to the following positions, see Fig.~\ref{fig:illustration-soliton-propagation-modes}(b):
\begin{list}{--}{\setlength{\leftmargin}{1.8em}\setlength{\rightmargin}{0em}
\setlength{\itemsep}{0ex}\setlength{\topsep}{0.5ex}}

\item[$y_1$:] position of the $(\nb+\nc)$-cluster, when the $\nc$-cluster detaches
 [first line in Fig.~\ref{fig:illustration-soliton-propagation-modes}(b)],

\item[$y_2$:] position of the $\nc$-cluster when it attaches to 
an $\nb$-cluster [third line in Fig.~\ref{fig:illustration-soliton-propagation-modes}(b)],

\item[$y_3$:] position of the $\nb$-cluster, when it attaches
to the $(\nc+\nb)$-cluster [fourth line in Fig.~\ref{fig:illustration-soliton-propagation-modes}(b)],

\item[$y_4$:] position of the $(\nb+\nc+\nb)$-cluster, when the composite $(\nc+\nb)$-cluster detaches 
[fifth line in Fig.~\ref{fig:illustration-soliton-propagation-modes}(b)].

\end{list}
The positions $y_1$ and $y_4$ for the detachment events are given by the requirement that the nonsplitting conditions for the
respective clusters at the respective positions are violated. This means that
$y_1$ is given by Eq.~\eqref{eq:y1-determ} and $y_4$ by
\begin{equation}
\frac{1}{\nb}\sum\limits_{j=0}^{\nb-1}\hspace{-0.3em}F(y_4+j\sigma)=
\frac{1}{\nb\!+\!\nc}\hspace{-0.3em}\sum\limits_{j=\nb}^{2\nb+\nc-1}\hspace{-1em}F(y_4+j\sigma)\,.
\label{eq:y4-determ}
\end{equation}

The position $y_3$ follows from
the requirement that the time $\tau_{\nb}(y_1,y_3)$
for the $\nb$-cluster to move from  $y_1$ to $y_3$ is equal to
the sum of times $\tau_{\nc}(y_1+\nb\sigma,y_2)$ and $\tau_{\nc+\nb}(y_2,y_3+\nb\sigma)$
for the $\nc$- and $(\ncb)$-clusters to move until the $\nb$ attaches to the $(\ncb)$-cluster:
\begin{equation}
\tau_{\nb}(y_1,y_3)=\tau_{\nc}(y_1+\nb\sigma,y_2)+\tau_{\nc+\nb}(y_2,y_3+\nb\sigma)\,.
\label{eq:y3-determ}
\end{equation}

If the time for the relaxation of the $\nb$-clusters 
is neglected,
$y_2$ is given by Eq.~\eqref{eq:y2inf} and the soliton velocity of the B type soliton becomes
\begin{equation}
v_{\rm sol}^{\rm B,\infty}=\frac{\mb}{\tau_{\rm sol}^{\rm B}(y,y_2^\infty,y_3^\infty,y_4)}\,,
\label{eq:vsolBinfty}
\end{equation}
where $y_3^\infty$ follows from Eq.~\eqref{eq:y3-determ} when $y_2^\infty$ is inserted into this equation.

When the relaxation of the $\nb$-clusters is taken into account, we again consider
the configuration, where the
soliton $\nc$-cluster attaches to the relaxing $\nb$-cluster initially at position $y_1$ after nearly one circulation around the 
system. As before, the soliton $\nc$-cluster and the $\nb$-cluster have positions $y_2'=y_2-\mbc$ and
$y_2'+\nc\sigma=y_2-\mbc+\nc\sigma$ at this moment. The time passed between the considered initial configuration
($\nb$-cluster at position $y_1$)
and the considered final configuration ($\nb$-cluster at position $y_2-\mbc+\nc\sigma$) is equal
to the time of the initially detaching $\nc$-cluster to move from $y_1+\nb\sigma$ to $y_2$ plus the time 
$(L-\mbc)/v_{\rm sol}=(L-\mbc\tau_{\rm sol}/\mb$ 
for the soliton to move the distance $L-\mbc$. The equation analogous to Eq.~\eqref{eq:y2-determ} thus
becomes
\begin{align}
&\tau_{\nb}(y_1,y_3)+\tau_{2\nb+\nc}(y_3,y_4)\nonumber\\
&\hspace{4em}{}+\tau_{\nb}(y_4,y_2-\mbc+\nc\sigma)
\label{eq:y2-determB}\\
&\hspace{0em}=\tau_{\nc}(y_1\!+\!\nb\sigma,y_2)+\frac{L\!-\!\mbc}{\mb}\,\tau_{\rm sol}^{\rm B}(y_1,y_2,y_3,y_4).
\nonumber
\end{align}
Inserting $\tau_{\rm sol}^{\rm B}(y_1,y_2,y_3,y_4)$ from Eq.~\eqref{eq:tausolB}, this Eq.~\eqref{eq:y2-determB} together
with Eq.~\eqref{eq:y3-determ} become two coupled determining equations for $y_2$ and $y_3$ whose solutions depend on $L$,
$y_2=y_2(L)$ and $y_3=y_3(L)$. 

Inserting these solutions into $\tau_{\rm sol}$ in Eq.~\eqref{eq:tausolB}
gives the relaxation-corrected soliton period, from which we obtain the relaxation-corrected soliton velocity
\begin{equation}
v_{\rm sol}^{\rm B}(1,L)= \frac{\mb}{\tau_{\rm sol}(y_1, y_2(L), y_3(L), y_4)}\,.
\label{eq:vsolB}
\end{equation}

Equation~\eqref{eq:vsol-multiple-solitons} remains unchanged for type B solitons, 
$v_{\rm sol}^{\rm B}(N_{\rm sol},L)=v_{\rm sol}^{\rm B}(1,L/N_{\rm sol})$.

%%%%%%%%%%%%%%%%%%%%%%%%%%%%%%%%%%%%%%%%%%%%%%%
\begin{figure}[t!]
\centering
\hspace*{-1.5em}
\includegraphics[width=0.8\columnwidth]{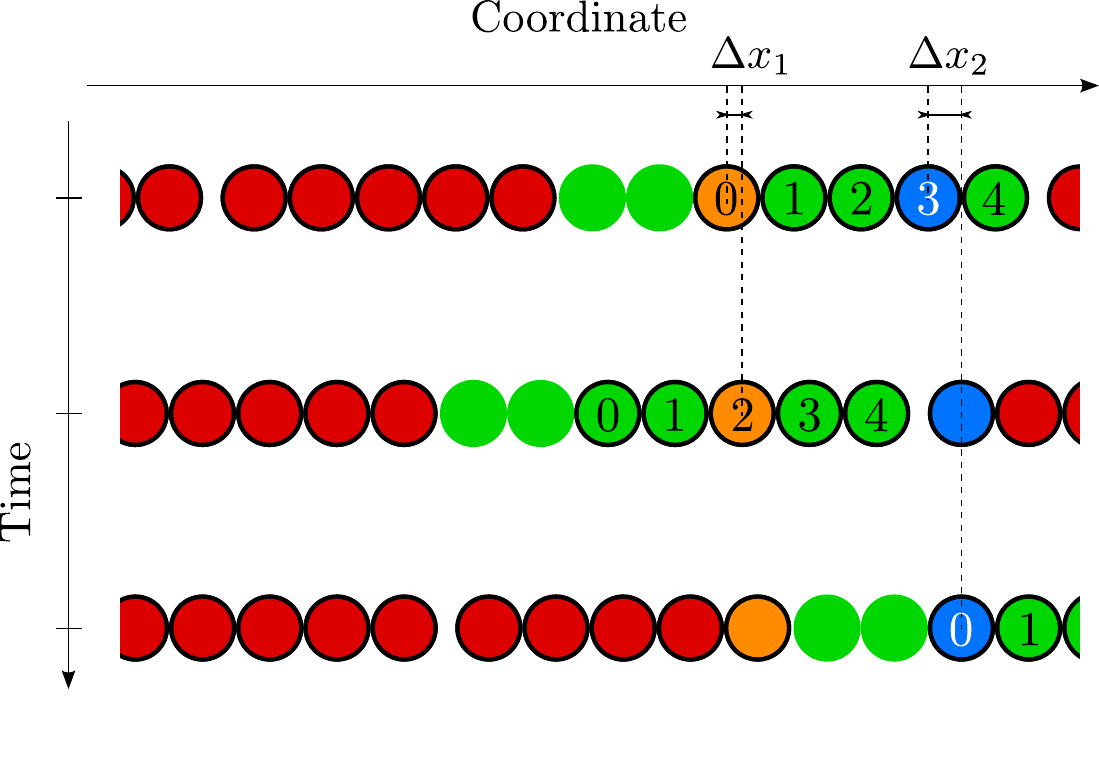}
\caption{Simulated particle configurations for the same parameters $\sigma=0.57$, $f=0.01$ and $L=40$ as in 
Fig.~\ref{fig:illustration-soliton-propagation-modes}(a) to 
illustrate the derivation of the soliton-mediated particle current ($N_{\rm sol}=1$, $\nb=5$, $\nc=2$).
In the first line, the $\nc$-cluster attaches to an $\nb$-cluster, where we have tagged the first particle (order number
$s=0$, orange) and
the fourth particle ($s=3$, blue). The second and third lines show the configurations after one soliton
circulation for the orange and blue tagged particles, respectively. 
After the soliton circulation, the orange tagged particle has order number $s'=s+\nc$
[change $\Delta s=\nc\ourmod\nb=2$, Eq.~\eqref{eq:s1}], and the blue tagged particle 
has order number $s'=0$ [change $\Delta s=\nc\ourmod\nb-\nb=\nc-\nb=-3$, Eq.~\eqref{eq:s2}].
Vertical dotted lines indicate the displacements $\Delta x_1$ and $\Delta x_2$ of the tagged particles corresponding to their
different type of order number change, see Eq.~\eqref{eq:Deltaxalpha}.}
\label{fig:illustration_for_formula_particle_current}
\end{figure}
%%%%%%%%%%%%%%%%%%%%%%%%%%%%%%%%%%%%%%%%%%%%%%%

\section{Displacement $d$ in Eq.~\eqref{eq:d}}
\label{app:d-displacement}

For deriving $d$, we need to specify positions of the soliton.
As initial particle configuration we consider one, where the
soliton $\nc$-cluster attaches to an $\nb$-cluster, see 
the first line in Fig.~B.1. In the respective $\nb$-cluster,
we tag a particle and calculate how it is displaced after successive soliton circulations. 
Each circulation is finished when the
soliton $\nc$-cluster attaches to the $\nb$-cluster that contains the tagged particle. Summing up all displacements 
of the tagged particle in the $k$ soliton circulations, with $k$ from Eq.~\eqref{eq:k-circulations}, we obtain $d$.

Let $s$ be the order number of the tagged particle and $y$ the position of the soliton ($\nc$-cluster) after the $j$th soliton circulation,
$j=0,\ldots,k-1$, see Fig.~B.1.
The position of the tagged particle in this configuration is
\begin{equation}
x=y+\nc\sigma+s\sigma\,.
\end{equation}
After one soliton circulation, the position of the soliton changes to $y'$, 
and the position and order number of the tagged particle to $x'$ and $s'$,
$x'=y'+\nc\sigma+s'\sigma$. Accordingly, the tagged particle is displaced by
\begin{equation}
\Delta x=x'-x=\Delta y+\Delta s\,\sigma\,,
\label{eq:Deltax}
\end{equation}
where $\Delta y=y'-y$ and $\Delta s=s'-s$. 

To derive the change $\Delta y$ of the soliton position, we note that
after each time period $\tau_{\rm sol}$, the soliton is displaced by $\mb$. 
Hence, the distance 
$\Delta Y$ traveled by the $\nc$-cluster in one circulation is
an integer multiple of $\mb$, 
$\Delta Y=m\mb$, 
$m\in\mathbb{N}$. Since $m$ must be larger
than the number $N_{\rm b}$ of $\nb$-clusters, we can write 
$m\hspace{0.05em}\nb=N_{\rm b}\nb+l\hspace{0.05em}\nb$,
where $l$ is a positive integer. The
change $\Delta y$ of the soliton position is equal to its distance traveled
modulo $L$, $\Delta y=\Delta Y\ourmod L=m\mb\ourmod L$.
Using $N_{\rm b}\mb+N_{\rm sol}\mbc=L$ from Eq.~\eqref{eq:filling} with $N_{\rm sol}=1$, we thus obtain 
\begin{equation}
\Delta y=l\mb-\mbc\,.
\label{eq:Deltay}
\end{equation}

Equations~\eqref{eq:Deltax}, \eqref{eq:Deltay}, \eqref{eq:s1} and \eqref{eq:s2} imply 
that only two displacements $\Delta x_{1,2}$ of the tagged particle can occur
after a soliton circulation:
\begin{equation}
\Delta x_\alpha=l_\alpha\mb-\mbc+\Delta s_\alpha\sigma\,,\hspace{1em}\alpha=1,2\,.
\label{eq:Deltaxalpha}
\end{equation}
Because $\Delta x_\alpha$ is larger than zero and smaller than $\mb$, $l_\alpha$ is the smallest integer
giving $\Delta x_\alpha>0$, which yields Eq.~\eqref{eq:lalpha} in the main text.

We finally have to sum up the $k$ displacements after each soliton circulation. Let $k_\alpha$ be the number 
of displacements $\Delta x_\alpha$. Then
\begin{gather}
k_1+k_2=k\,,\\
k_1\Delta s_1+k_2\Delta s_2=0\,,
\end{gather}
where the second equation follows from the fact that the total change of the tagged particle's order number is zero
after the $k$ soliton circulations. Solving for the $k_\alpha$ and inserting the results from Eqs.~\eqref{eq:s1}, \eqref{eq:s2}
and \eqref{eq:k-circulations}, we obtain
\begin{align}
k_1&=-\frac{\Delta s_2 k}{\Delta s_1-\Delta s_2}=\frac{\nb-\nc\ourmod\nb}{\gcd(\nb,\nc)}\,,\\
k_2&=\frac{\Delta s_1 k}{\Delta s_1-\Delta s_2}=\frac{\nc\ourmod\nb}{\gcd(\nb,\nc)}\,.
\end{align}
Note that $(\nc\ourmod\nb)$ is divisible by $\gcd(\nb,\nc)$. 
The total displacement $d$ after the $k$ soliton circulations is
\begin{align}
d&=k_1\Delta x_1+k_2\Delta x_2\nonumber\\
&=\frac{\nb\!-\!\nc\ourmod\nb}{\gcd(\nb,\nc)}l_1\mb+\frac{\nc\ourmod\nb}{\gcd(\nb,\nc)}l_2\mb\nonumber\\
&\hspace{1em}{}-k\mbc\,,
\end{align}
which is the result given in Eq.~\eqref{eq:d} of the main text.

%\bibliographystyle{elsarticle-num} 
%\bibliography{driven-systems-local-short.bib}

\begin{thebibliography}{10}
\expandafter\ifx\csname url\endcsname\relax
  \def\url#1{\texttt{#1}}\fi
\expandafter\ifx\csname urlprefix\endcsname\relax\def\urlprefix{URL }\fi
\expandafter\ifx\csname href\endcsname\relax
  \def\href#1#2{#2} \def\path#1{#1}\fi

\bibitem{Xiao/etal:2003}
W.~Xiao, P.~A. Greaney, D.~C. Chrzan, Adatom transport on strained {C}u(001):
  Surface crowdions, Phys. Rev. Lett. 90 (2003) 156102.

\bibitem{Paneth:1950}
H.~R. Paneth, The mechanism of self-diffusion in alkali metals, Phys. Rev. 80
  (1950) 708--711.

\bibitem{Zepeda-Ruiz/etal:2004}
L.~A. Zepeda-Ruiz, J.~Rottler, S.~Han, G.~J. Ackland, R.~Car, D.~J. Srolovitz,
  Strongly non-{A}rrhenius self-interstitial diffusion in vanadium, Phys. Rev.
  B 70 (2004) 060102.

\bibitem{Korznikova/etal:2022}
E.~Korznikova, V.~Shunaev, I.~Shepelev, O.~Glukhova, S.~Dmitriev, Ab initio
  study of the propagation of a supersonic 2-crowdion in fcc al, Comput. Mat.
  Sci. 204 (2022) 111125.

\bibitem{Landau/etal:1993}
A.~I. Landau, A.~S. Kovalev, A.~D. Kondratyuk, Model of interacting atomic
  chains and its application to the description of the crowdion in an
  anisotropic crystal, Phys. Status Solidi B 179~(2) (1993) 373--381.

\bibitem{Matsukawa/Zinkle:2007}
Y.~Matsukawa, S.~J. Zinkle, One-dimensional fast migration of vacancy clusters
  in metals, Science 318~(5852) (2007) 959--962.

\bibitem{vanderMeer/etal:2018}
B.~van~der Meer, R.~van Damme, M.~Dijkstra, F.~Smallenburg, L.~Filion,
  Revealing a vacancy analog of the crowdion interstitial in simple cubic
  crystals, Phys. Rev. Lett. 121 (2018) 258001.

\bibitem{Korda/etal:2002}
P.~T. Korda, M.~B. Taylor, D.~G. Grier, Kinetically locked-in colloidal
  transport in an array of optical tweezers, Phys. Rev. Lett. 89 (2002) 128301.

\bibitem{Bohlein/etal:2012}
T.~Bohlein, J.~Mikhael, C.~Bechinger, Observation of kinks and antikinks in
  colloidal monolayers driven across ordered surfaces, Nat. Mater. 11~(2)
  (2012) 126--130.

\bibitem{Bohlein/Bechinger:2012}
T.~Bohlein, C.~Bechinger, Experimental observation of directional locking and
  dynamical ordering of colloidal monolayers driven across quasiperiodic
  substrates, Phys. Rev. Lett. 109 (2012) 058301.

\bibitem{Vanossi/etal:2012}
A.~Vanossi, N.~Manini, E.~Tosatti, Static and dynamic friction in sliding
  colloidal monolayers, Proc. Natl. Acad. Sci. U.S.A. 109~(41) (2012)
  16429--16433.

\bibitem{Brazda/etal:2018}
T.~Brazda, A.~Silva, N.~Manini, A.~Vanossi, R.~Guerra, E.~Tosatti,
  C.~Bechinger, Experimental observation of the {Aubry} transition in
  two-dimensional colloidal monolayers, Phys. Rev. X 8 (2018) 011050.

\bibitem{Vanossi/etal:2020}
A.~Vanossi, C.~Bechinger, M.~Urbakh, Structural lubricity in soft and hard
  matter systems, Nat. Commun. 11~(1) (2020) 4657.

\bibitem{deSouzaSilva/etal:2006}
C.~C. de~Souza~Silva, J.~Van~de Vondel, M.~Morelle, V.~V. Moshchalkov,
  Controlled multiple reversals of a ratchet effect, Nature 440~(7084) (2006)
  651--654.

\bibitem{Tierno/Fischer:2014}
P.~Tierno, T.~M. Fischer, Excluded volume causes integer and fractional
  plateaus in colloidal ratchet currents, Phys. Rev. Lett. 112 (2014) 048302.

\bibitem{Stoop/etal:2020}
R.~L. Stoop, A.~V. Straube, T.~H. Johansen, P.~Tierno, Collective directional
  locking of colloidal monolayers on a periodic substrate, Phys. Rev. Lett. 124
  (2020) 058002.

\bibitem{Lips/etal:2021}
D.~Lips, R.~L. Stoop, P.~Maass, P.~Tierno, Emergent colloidal currents across
  ordered and disordered landscapes, Commun. Phys. 4~(1) (2021) 224.

\bibitem{Juniper/etal:2015}
M.~P.~N. Juniper, A.~V. Straube, R.~Besseling, D.~G. A.~L. Aarts, R.~P.~A.
  Dullens, Microscopic dynamics of synchronization in driven colloids, Nat.
  Commun. 6~(1) (2015) 7187.

\bibitem{Vanossi/Tosatti:2012}
A.~Vanossi, E.~Tosatti, Kinks in motion, Nat. Mater. 11~(2) (2012) 97--98.

\bibitem{Braun/Kivshar:2004}
O.~M. Braun, Y.~S. Kivshar, The Frenkel-Kontorova Model: Concepts, Methods, and
  Applications, Springer Berlin, Heidelberg, 2004.

\bibitem{Antonov/etal:2022a}
A.~P. Antonov, A.~Ryabov, P.~Maass, Solitons in overdamped {B}rownian dynamics,
  Phys. Rev. Lett. 129 (2022) 080601.

\bibitem{Cereceda-Lopez/etal:2023}
E.~Cereceda-L{\'o}pez, A.~P. Antonov, A.~Ryabov, P.~Maass, P.~Tierno,
  Overcrowding induces fast colloidal solitons in a slowly rotating potential
  landscape, Nat. Commun. 14~(1) (2023) 6448.

\bibitem{suppl-solitary-cluster-waves}
See Supplemental Material for movies of solitary cluster wave propagation.

\bibitem{Ablowitz:1991}
M.~A. Ablowitz, P.~A. Clarkson, Solitons, Nonlinear Evolution Equations and
  Inverse Scattering, London Mathematical Society Lecture Note Series,
  Cambridge University Press, Cambridge, 1991.

\bibitem{Malomed/etal:2020}
B.~A. Malomed, Nonlinearity and Discreteness: Solitons in Lattices, Springer
  International Publishing, Cham, 2020, pp. 81--110.

\bibitem{Strecker/etal:2002}
K.~E. Strecker, G.~B. Partridge, A.~G. Truscott, R.~G. Hulet, Formation and
  propagation of matter-wave soliton trains, Nature 417~(6885) (2002) 150--153.

\bibitem{Carr/Brand:2004}
L.~D. Carr, J.~Brand, Spontaneous soliton formation and modulational
  instability in {B}ose-{E}instein condensates, Phys. Rev. Lett. 92 (2004)
  040401.

\bibitem{Karpov/etal:2016}
M.~Karpov, H.~Guo, A.~Kordts, V.~Brasch, M.~H.~P. Pfeiffer, M.~Zervas,
  M.~Geiselmann, T.~J. Kippenberg, Raman self-frequency shift of dissipative
  {K}err solitons in an optical microresonator, Phys. Rev. Lett. 116 (2016)
  103902.

\bibitem{Kippenberg/etal:2018}
T.~J. Kippenberg, A.~L. Gaeta, M.~Lipson, M.~L. Gorodetsky, Dissipative {K}err
  solitons in optical microresonators, Science 361~(6402) (2018) eaan8083.

\bibitem{Deconinck/etal:1993}
B.~Deconinck, P.~Meuris, F.~Verheest, Oblique nonlinear {A}lfv{\'e}n waves in
  strongly magnetized beam plasmas. {P}art 2. {S}oliton solutions and
  integrability, J. Plasma Phys. 50~(3) (1993) 457--476.

\bibitem{Xu/etal:2008}
T.~Xu, B.~Tian, L.-L. Li, X.~L{\"u}, C.~Zhang, {Dynamics of Alfv{\'e}n solitons
  in inhomogeneous plasmas}, Phys. Plasmas 15~(10) (2008) 102307.

\bibitem{Malomed:1992}
B.~A. Malomed, Propagating solitons in damped ac-driven chains, Phys. Rev. A 45
  (1992) 4097--4101.

\bibitem{Kuusela:1992}
T.~Kuusela, Soliton experiments in a damped ac-driven nonlinear electrical
  transmission line, Phys. Lett. A 167~(1) (1992) 54--59.

\bibitem{Kuusela/etal:1993}
T.~Kuusela, J.~Hietarinta, B.~A. Malomed, Numerical study of solitons in the
  damped ac-driven {T}oda lattice, J. Phys. A: Math. Gen. 26~(1) (1993) L21.

\bibitem{Hietarinta/etal:1995}
J.~Hietarinta, T.~Kuusela, B.~A. Malomed, Shock waves in the dissipative {T}oda
  lattice, J. Phys. A Math. Gen. 28~(11) (1995) 3015.

\bibitem{Malomed/etal:1990}
B.~A. Malomed, V.~A. Oboznov, A.~V. Ustinov, ``{S}upersolitons'' in
  periodically inhomogeneous long {J}osephson junctions, Zh. Eksp. Teor. Fiz.
  97 (1990) 924.

\bibitem{Malomed:1990}
B.~A. Malomed, Superfluxons in periodically inhomogeneous long {J}osephson
  junctions, Phys. Rev. B 41 (1990) 2616--2618.

\bibitem{Madsen/etal:2008}
P.~A. Madsen, D.~R. Fuhrman, H.~A. Sch{\"a}ffer, On the solitary wave paradigm
  for tsunamis, J. Geophys. Res.: Oceans 113~(C12) (2008) C12012.

\bibitem{Constantin/Henry:2009}
A.~Constantin, D.~Henry, Solitons and tsunamis, Z. Naturforsch. A 64~(1-2)
  (2009) 65--68.

\bibitem{Stanton/Ostrovsky:1998}
T.~P. Stanton, L.~A. Ostrovsky, Observations of highly nonlinear internal
  solitons over the continental shelf, Geophys. Res. Lett. 25~(14) (1998)
  2695--2698.

\bibitem{Risken:1985}
H.~Risken, The {F}okker-{P}lanck {E}quation: {M}ethods of {S}olution and
  {A}pplications, Springer-Verlag Berlin, 1985.

\bibitem{Antonov/etal:2022c}
A.~P. Antonov, S.~Schweers, A.~Ryabov, P.~Maass, Brownian dynamics simulations
  of hard rods in external fields and with contact interactions, Phys. Rev. E
  106 (2022) 054606.

\bibitem{Vorobyov:1980}
N.~N. Vorobyov, Criteria for divisibility, University of Chicago Press, 1980.

\bibitem{Abramowitz/Stegun:1965}
M.~Abramowitz, I.~Stegun, Handbook of Mathematical Functions: With Formulas,
  Graphs, and Mathematical Tables, Applied mathematics series, Dover
  Publications, 1965.

\bibitem{Antonov:2023}
A.~Antonov, Brownian particle transport in periodic structures, Ph.D. thesis,
  Osnabr\"uck University, Germany (Aug 2023).

\bibitem{Lips/etal:2018}
D.~Lips, A.~Ryabov, P.~Maass, Brownian asymmetric simple exclusion process,
  Phys. Rev. Lett. 121 (2018) 160601.

\bibitem{Lips/etal:2019}
D.~Lips, A.~Ryabov, P.~Maass, Single-file transport in periodic potentials:
  {The Brownian} asymmetric simple exclusion process, Phys. Rev. E 100 (2019)
  052121.

\bibitem{teVrugt/etal:2020}
M.~te~Vrugt, H.~L{\"o}wen, R.~Wittkowski, Classical dynamical density
  functional theory: from fundamentals to applications, Adv. Phys. 69~(2)
  (2020) 121--247.

\bibitem{Schmidt:2022}
M.~Schmidt, Power functional theory for many-body dynamics, Rev. Mod. Phys. 94
  (2022) 015007.

\bibitem{delasHeras/etal:2023}
D.~de~las Heras, T.~Zimmermann, F.~Samm{\"u}ller, S.~Hermann, M.~Schmidt,
  Perspective: How to overcome dynamical density functional theory, J. Phys.
  Condens. Matter 35~(27) (2023) 271501.

\bibitem{Bertini/etal:2015}
L.~Bertini, A.~De~Sole, D.~Gabrielli, G.~Jona-Lasinio, C.~Landim, Macroscopic
  fluctuation theory, Rev. Mod. Phys. 87 (2015) 593--636.

\bibitem{Lutz/etal:2006}
C.~Lutz, M.~Reichert, H.~Stark, C.~Bechinger, Surmounting barriers: {{The}}
  benefit of hydrodynamic interactions, EPL 74~(4) (2006) 719--725.

\bibitem{Tsuji/etal:2020}
T.~Tsuji, R.~Nakatsuka, K.~Nakajima, K.~Doi, S.~Kawano, Effect of hydrodynamic
  inter-particle interaction on the orbital motion of dielectric nanoparticles
  driven by an optical vortex, Nanoscale 12 (2020) 6673--6690.

\bibitem{Tsuji/etal:2022}
T.~Tsuji, K.~Doi, S.~Kawano, Optical trapping in micro- and nanoconfinement
  systems: Role of thermo-fluid dynamics and applications, J. Photochem.
  Photobiol. C: Photochem. Rev. 52 (2022) 100533.

\bibitem{Antonov/etal:2022b}
A.~P. Antonov, D.~Vor{\'{a}}{\v{c}}, A.~Ryabov, P.~Maass, Collective
  excitations in jammed states: ultrafast defect propagation and finite-size
  scaling, New J. Phys. 24~(9) (2022) 093020.

\bibitem{Stuhlmueller/etal:2022}
N.~C.~X. Stuhlm\"uller, T.~M. Fischer, D.~de~las Heras, Colloidal transport in
  twisted lattices of optical tweezers, Phys. Rev. E 106 (2022) 034601.

\bibitem{Stuhlmueller/etal:2024}
N.~C.~X. Stuhlm{\"u}ller, T.~M. Fischer, D.~de~las Heras, Competition between
  drift and topological transport of colloidal particles in twisted magnetic
  patterns, New J. Phys. 26 (2024) 023056.

\end{thebibliography}

\end{document}